\documentclass[a4paper,11pt]{article}
\pdfoutput=1 
\usepackage{amssymb}
\usepackage{siunitx}
\usepackage{amsmath}
\usepackage{braket}
\usepackage{mathtools}
\usepackage[usenames,dvipsnames,svgnames,table]{xcolor}
\usepackage [utf8] {inputenc}
\usepackage{color}
\usepackage{subcaption}
\usepackage{lmodern}
\usepackage{footnote}
\usepackage{slashed}
 \usepackage{mathrsfs}
 \usepackage[pdftex] {graphicx}

\usepackage{multirow}
\usepackage{jheppub} 
\usepackage{here}
\usepackage{hyperref}
\usepackage[justification=centering, singlelinecheck=false]{caption}
\usepackage[list=true, labelfont=bf, labelformat=brace, position=top]{subcaption}

\newcommand{\eq}{\begin{eqnarray}}
\newcommand{\en}{\end{eqnarray}}
\newcommand{\gnabla}{{\stackrel{\leftrightarrow}{\nabla}}}

\preprint{MIT-CTP/5433, DESY-22-080 }
\title{Resonance form factors from finite-volume correlation functions
  with the external field method}
\author[1]{Jonathan Lozano,}
\affiliation[1]{Helmholtz-Institut f\"ur Strahlen- und Kernphysik (Theorie) and\\ Bethe Center for Theoretical Physics, Universit\"at Bonn, 53115 Bonn, Germany}

\emailAdd{lozano@hiskp.uni-bonn.de}

\author[1,2,3]{Ulf-G. Mei{\ss}ner,}

\affiliation[2]{Institute for Advanced Simulation and Institut
  f{\"u}r Kernphysik, Forschungszentrum J\"ulich,\\ D-52425
  J\"ulich, Germany}

\affiliation[3]{Tbilisi State  University,  0186 Tbilisi, Georgia}

\emailAdd{meissner@hiskp.uni-bonn.de}

\author[4]{Fernando Romero-L\'opez,}

\affiliation[4]{Center  for  Theoretical  Physics,  Massachusetts
  Institute  of  Technology,\\  Cambridge,  MA  02139,  USA}

\emailAdd{fernando@mit.edu}

\author[1,3]{Akaki Rusetsky,}
\emailAdd{rusetsky@hiskp.uni-bonn.de}

\author[5]{and Gerrit Schierholz}

\affiliation[5]{Deutsches Elektronen-Synchrotron DESY,
  Notkestr. 85, 22607 Hamburg, Germany}

\emailAdd{gerrit.schierholz@desy.de}

\abstract{
    \noindent
        A novel method for the extraction of form factors of unstable
    particles on the lattice is proposed. The approach is based on the study of two-particle
    scattering in a static, spatially periodic external field by using
    a generalization of the L\"uscher method in the presence
    of such a field. It is shown that
    the resonance form factor is given
    by the derivative of the resonance pole position in the complex plane
    with respect to the coupling constant to the external field.
    Unlike the standard approach, this proposal does not suffer from problems     
    caused by the presence of the triangle diagram.        
}

\allowdisplaybreaks
\begin{document}
\maketitle
\flushbottom
\clearpage
\section{Introduction}

The study of the form factors of unstable particles from lattice field theory provides plenty of
information about the structure of these particles. This study is however complicated
by a non-trivial mapping of the results of lattice calculations --- performed
in a finite volume --- onto the relevant infinite-volume form factors\footnote{Discretization effects are neglected throughout this paper.}.

Such a mapping is rather trivial in case of a stable particle. Namely, let $|p\rangle$ be an infinite-volume state, which describes a single particle moving with the on-shell momentum
$p^\mu$. The infinite-volume form factor
$\langle p|J(0)|q\rangle$ is defined as the matrix element of some current $J(x)$
between one-particle states (in order to ease the notations, we consider the spinless
particles and the scalar currents here). Due to the Lorentz-invariance, the form factor is a function of a single variable
$t=(p-q)^2$. Furthermore, on the lattice,
the spectrum contains the one-particle states $|p\rangle_L$. Here, $L$ denotes
the spatial extension of a cubic lattice, while, for simplicity,
the time extension is assumed to be infinite. The finite-volume form factor is given by the matrix
element between these states, $\langle p|J(0)|q\rangle_L$. Then, in the limit $L\to\infty$, one has
\eq\label{eq:L-infinity}
\langle p|J(0)|q\rangle_L=\langle p|J(0)|q\rangle+O(e^{-\mu L})\, ,
\en
where $\mu$ is a characteristic mass scale---typically a multiple of the mass of lightest particle
in the system. Furthermore, recall that three-momenta in a finite volume are discretized,
${{\bf p}=(2\pi/L)\,{\bf n}}$ and ${\bf n}\in\mathbb{Z}^3$.
Hence, in order to have a fixed ${\bf p}$ in infinite volume,
one cannot keep ${\bf n}$ constant. One could, for example,
choose a monotonic sequence of discrete values of $L \in \{L_i\}$, such that ${\bf p}$ and ${\bf q}$ are allowed finite-volume momenta.
Equation (\ref{eq:L-infinity}) must be interpreted exactly in this sense.

The situation is far less trivial in case of unstable particles. First, a one-particle
state describing a resonance does not exist in the infinite-volume spectrum. Let us, for
simplicity, consider a situation in which the resonance emerges in the scattering of two
identical spinless particles. In order to define the resonance form factor in the infinite
volume, one has to start from the five-point Green function
$\langle p_1,p_2;out|J(0)|q_1,q_2;in\rangle$. Defining the total momenta of the
outgoing and incoming particle pair by $P=p_1+p_2$ and $Q=q_1+q_2$, respectively, it can
be shown that, if a resonance in a given channel exists,
this five-point function possesses a double pole in the complex plane,
\begin{equation}
\langle p_1,p_2;out|J(0)|q_1,q_2;in\rangle  \sim\dfrac{1}{(M_R^2-P^2)(M_R^2-Q^2)},
\end{equation}
located on some unphysical Riemann sheet for the
variables $P^2,Q^2$. The infinite-volume
resonance form factor can be expressed through the residue at this double pole (for more details,
see e.g, Ref.~\cite{Hoja:2010fm}). This (complex-valued)
form factor is a function of a single variable
$t=(P-Q)^2$ and, in case of the conserved currents, obeys  the usual Ward identities
--- for example, it is properly normalized at $t=0$.

In a finite volume, one can access
the spectrum of a Hamiltonian having the quantum number of two particle states.
Let us denote the eigenstates of the Hamiltonian by $|\alpha,{\bf P}\rangle_L$ (the so-called scattering states). Here, ${\bf P}$
is the total three-momentum of two particles and $\alpha$ labels different states
having the same ${\bf P}$. If one varies
$L$ while keeping the ${\bf P}$ constant, the energies $E_\alpha({\bf P},L)$
exhibit power-law corrections in $L$ with respect to the sum of the energies of one-particle states. Furthermore,  one can evaluate the matrix elements of a current
$\langle \alpha,{\bf P}|J(0)|\beta,{\bf Q}\rangle_L$ on the lattice for any $\alpha,\beta$. Interpreting an infinite-volume
limit of such a matrix element, as well as performing the analytic continuation to the resonance pole is however a delicate task. As in case of a stable particle, the momenta
${\bf P},{\bf Q}$ are discretized and the limit $L\to\infty$ has to be treated accordingly
(namely, the pertinent integer vectors ${\bf n},{\bf m}$ cannot be considered fixed, see the discussion above).
Furthermore, even for a fixed ${\bf P}$, the eigenvalues $E_\alpha({\bf P},L)$
collapse toward the threshold, as $L\to\infty$. Therefore, in order
to stay in the vicinity of a fixed infinite-volume center-of-mass (CM) energy $E$
(or, equivalently, $E_\alpha({\bf P},L)\simeq \sqrt{E^2+{\bf P}^2}$), one cannot treat $\alpha$
as fixed anymore\footnote{Note that in Ref.~\cite{Briceno:2020xxs}, the finite-volume matrix elements at a fixed $\alpha,\beta$  are considered. For instance, the ground-state matrix element can be expanded in $1/L$, which gives only access to the form factor at zero-momentum transfer. By contrast, matrix elements at fixed energy have irregular behavior as a function of $L$, “jumping” over the poles of the one-loop diagrams in a finite volume.
}.
Higher excited states should be considered in the limit $L\to\infty$ and fixed $E$.

After fixing carefully the kinematics, one might ask oneself, how the
infinite-volume limit should be carried out in the matrix elements. For instance,
it is well known that the corrections are no more exponentially suppressed for
unstable particles. Even for a much simpler case of the finite-volume
decay matrix element of
an unstable state, this limit is not well defined mathematically and can be performed
only after
removing the factor that corresponds to the interactions of the decay
products in the final state, the well-known the Lellouch-L\"uscher
factor~\cite{Lellouch:2000pv}. This approach works perfectly for
the transition form factors of resonances into stable states,
as well as for 
timelike form factors of stable particles, see~\cite{Hansen:2012tf,Briceno:2014uqa,Briceno:2015csa,Briceno:2021xlc,Briceno:2016kkp,Briceno:2015dca,Agadjanov:2016fbd,Agadjanov:2014kha,Meyer:2011um,Sherman:2022tco}.
Recently, a three-particle analog of the Lellouch-L\"uscher formula has been also
derived~\cite{Muller:2020wjo,Hansen:2021ofl}.
The situation, however, becomes more complicated in case of the resonance
matrix elements which is studied in the present paper.
The problem is that, even after explicitly removing
the Lellouch-L\"uscher factors that correspond to the
unstable particles, the remaining expression still does not exhibit a regular
behavior in $L$ and, hence, the infinite-volume limit cannot be
performed~\cite{Hoja:2010fm,Bernard:2012bi,Briceno:2015tza,Baroni:2018iau}. Additional developments concerning the evaluation of the resonance matrix elements can be found in Refs.~\cite{Briceno:2019nns,Briceno:2020vgp,Briceno:2020xxs}.

\begin{figure}[t]
  \begin{center}
    \includegraphics*[width=8.cm]{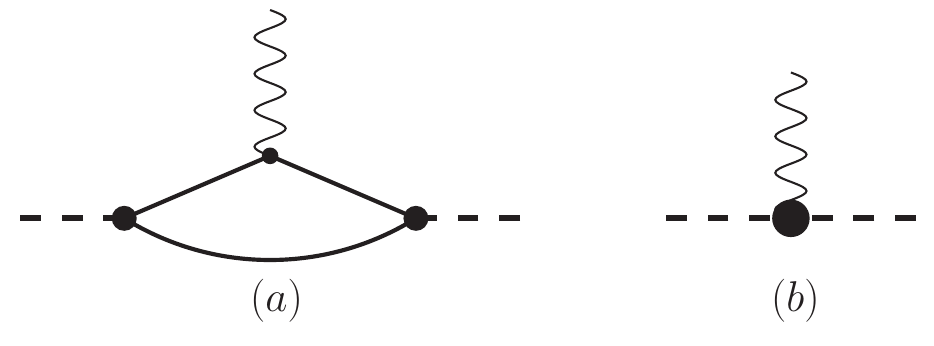}
    \caption{a) The triangle diagram which leads to an irregular behavior
      of the matrix element in a finite volume, and b) a local vertex that has a regular behavior.
      The dashed, single and wiggled lines denote the resonance, its constituents and the external current,
      respectively.}
    \label{fig:triangle}
  \end{center}
  \end{figure}

 To summarize the findings of Refs.~\cite{Hoja:2010fm,Bernard:2012bi,Briceno:2015tza,Baroni:2018iau}, a consistent procedure for the analytic continuation
of the obtained result into the complex plane, which is needed to define
a resonance form factor rigorously, cannot be straightforwardly formulated for the whole
finite-volume matrix element. The culprit
is the so-called triangle diagram, in which one of the
``constituent particles'' of a resonance couples to the external current $J$, whereas
the second acts as a spectator, see Fig.~\ref{fig:triangle}a
(for simplicity, we consider  the resonance emerging in two-particle scattering). 
Such a triangle diagram is more singular in the finite volume than a loop diagram with two propagators,
which corresponds to the L\"uscher zeta-function. In
Refs.~\cite{Hoja:2010fm,Bernard:2012bi,Briceno:2015tza,Baroni:2018iau}
the problem was addressed in different frameworks, but from a very similar physics
perspective. Schematically, the proper procedure could be described as follows.
It is proposed to single out the contribution of the triangle diagram in a finite
volume. The infinite-volume limit and the analytic continuation in the remainder of the
amplitude can be performed without further ado. The triangle contribution, calculated
analytically in the infinite volume and at the resonance pole, can be added back at the final stage.
Even if the above procedure is absolutely consistent,
the necessity of subtracting/adding  the triangle diagram, to our taste,
may turn the extraction of the resonance form factors into quite a challenging
endeavor, with hard-to-control systematic errors.

On the other hand, the Feynman-Hellmann theorem~\cite{Hellmann,Feynman:1939zza}
has been successfully used to compute form factors of {\em stable} hadrons
in a static, spatially periodic external field~\cite{QCDSF:2017ssq}\footnote{Note however that the study of the limit of zero-momentum transfer in this approach requires further scrutiny and is by no means trivial. The structure
of the energy levels changes in this limit --- the Landau levels emerge in the constant field. More discussion of this subtle issue is given in Ref.~\cite{Agadjanov:2018yxh}.}.
Moreover, the same method has been applied to the study of
baryon structure functions and doubly virtual Compton scattering amplitude on the
lattice~\cite{Can:2020sxc,Hannaford-Gunn:2021mrl,Agadjanov:2016cjc,Agadjanov:2018yxh}. In case of the
form factor, one computes the two-point function {\em in an external field} in the Breit
frame\footnote{Since the external field breaks translational invariance, the three-momentum is not conserved.
  The two-point function in the Breit frame is then defined as the one
  whose initial and final three-momenta ${\bf p},{\bf q}$
  are opposite in direction and have the same magnitude $|{\bf p}|=|{\bf q}|=\omega/2$, where $2\pi/\omega$
  defines the period of the external field.}, and determines the mass of a particle in the external
field. It can then be shown that the derivative of the particle mass with respect to the coupling
constant to the external field gives, at leading order in this coupling constant, the form factor
in the Breit frame.

It is natural to ask, whether one can generalize this method to the calculation of the form factors of
unstable particles. The role of the particle mass in this case is played by the resonance pole
position in the complex plane.
In this paper we shall demonstrate that, at leading order,
the derivative of the pole position with respect to the coupling constant
to the external field gives the resonance form factor.
It will also be shown that, in order to compute the resonance
form factor, performing analytic continuation and finding a pole
in the presence of the external
field is even superfluous. It suffices to determine the local contribution
to the form factor, see Fig.~\ref{fig:triangle}b, which can be extracted directly
on the real axis. Then, the analytic continuation can be performed in the
explicit expressions, evaluated in the infinite volume and in the absence of the external
field. This does not cause any problems and hence, the problem related to the triangle
diagram, does not show up in this approach. Note that the local contribution, unlike
the matrix element itself, contains only exponentially suppressed corrections in a finite
volume, which are easy to handle.
Finally, note
that applying the Feynman-Hellmann theorem to resonances is not new. In particular, this
theorem has been used to define the sigma-terms for the resonances in Ref.~\cite{RuizdeElvira:2017aet}.
In the present paper, the approach of Ref.~\cite{RuizdeElvira:2017aet} is generalized to the case of
spatially periodic fields\footnote{Here, we mention in addition an application of the
  Feynman-Hellmann theorem to the calculation of the matrix element of a current
  between the two-body scattering states, which was carried out
  in Ref.~\cite{Briceno:2020xxs}. However, as already mentioned, the results of that paper cannot be directly compared to ours. First, in our calculations, the CM energies of the incoming/outgoing pairs and not the labels $\alpha,\beta$ are fixed. Second, the Feynman-Hellmann theorem in Ref.~\cite{Briceno:2020xxs} is used for the energies on the real axis and not in the complex plane. Lastly, the method of Ref.~\cite{Briceno:2020xxs} is restricted to the scalar current which can be obtained through the differentiation of the Lagrangian over the particle mass, whereas the method, described in the present paper,
  can be applied to a generic current.}.

The material of the present paper is rather technical. In order to make the argument easier to
follow, here we present a brief synopsis with the focus on the physics content. 

\begin{itemize}

\item[i)]
In our derivation, we make use
of the framework of the non-relativistic effective field theory (NREFT).
As already mentioned, the contributions to the form factor fall into two classes. This is shown
  in Fig.~\ref{fig:triangle}, where the triangle diagram causes difficulties in the infinite-volume limit,
  and the local contribution does not. Note also that all ingredients needed to construct the
  triangle diagram are assumed to be known in advance: the external field coupling to a single
  constituent, and a resonance coupling to the constituents (described by the elastic phase shift
  at the energies close  to the resonance mass). By contrast, the contact contribution is unknown
  and should be determined. Such a splitting
  can be naturally described in an NREFT framework, with a Lagrangian similar
  to the one given below in Eq.~(\ref{eq:Lagrangian}). Here, the coupling $C_R,\ldots$
  determines the single particle form factor, the couplings $C_0,C_2$ describe
  the S-wave elastic scattering phase shift near threshold, and the quantity $\kappa$
  characterizes the lowest-order contact interaction. Higher-order terms are not displayed explicitly. At this order, determining $\kappa$ on the lattice is equivalent
to extracting the form factor, which can be straightforwardly calculated from the known analytic expression in the infinite volume that contains $\kappa$ (as well as other constants)
as free parameters.
  
\item[ii)]
  In order to describe the form factor, one has to inject a non-zero momentum transfer
  between the initial and final states. This can be achieved by placing a system in a spatially
  periodic external field, whose frequency is equal to the momentum transfer.
  The details are considered in Sect.~\ref{sec:Mathieu}, where the result of
  Ref.~\cite{QCDSF:2017ssq} concerning the determination of the form factor
  of a stable particle using the Feynman-Hellmann method has been re-derived
  and extended (see Sect.~\ref{sec:energyshift-1}).
\item[iii)]
  The central result of the paper is the derivation of the generalized L\"uscher equation
  in the periodic external field. Symbolically, this can be written as
  \eq
  \det\biggl(X^{-1}-\frac{1}{2}\,\Pi\biggr)=0\, ,
\en
see Eqs.~(\ref{eq:luescher_ext}) and (\ref{eq:Xm1}) for more details.
Here, the matrix $X^{-1}$ is a counterpart of the inverse $K$-matrix,  $p\cot\delta(p)$,
and the loop function $\Pi$ corresponds to the L\"uscher zeta-function
$Z_{00}$, when the external periodic field is turned on. 
This equation enables one to extract the contact contribution $\kappa$ from the
two-particle energy spectrum in the external field, provided the single particle
form factor and the phase shift have been computed in advance. The extraction of $\kappa$ can be
performed at real energies, an analytic continuation is not needed. The infinite-volume
limit is trivial since, by definition, the contact terms can contain only exponentially
suppressed corrections for the large box sizes. 
  
\item[iv)]
  The Feynman-Hellmann theorem in quantum mechanics deals with the Hamiltonians,
  linearly depending on a parameter $\lambda$:
  \eq
  H=H_0+\lambda{\cal O}\, ,
  \en
  where ${\cal O}$ is some operator. The eigenstates of the Hamiltonian,
  $|n(\lambda)\rangle$, and the eigenvalues, $E_n(\lambda)$, also depend
  on $\lambda$. The Feynman-Hellmann theorem states that
  \eq
  \frac{d E_n(\lambda)}{d\lambda}=\langle n(\lambda)|{\cal O}|n(\lambda)\rangle\, .
  \en
  We generalize this result for unstable states. One can namely
  extract the resonance pole position $P_R^0$
  in the complex energy plane, also for a
  non-vanishing external field, see Sect.~\ref{sec:extracting}.
The derivative of $P_R^0$ with respect to the coupling to the external
field, $e$ (which plays the role of $\lambda$ here),
at $e=0$ is proportional to the resonance form factor, evaluated
  at the (complex) resonance pole:
  \eq
  \frac{dP_R^0(e)}{de}\biggr|_{e=0}\propto F\, .
  \en
  For more details, see Sect.~\ref{sec:relation} and, in particular, Eq.~(\ref{eq:main}).

\end{itemize}  

The layout of the paper is as follows:
In Sect.~\ref{sec:infinite} we consider the problem exclusively in the infinite volume and give
a consistent definition of the resonance form factor.
Sect.~\ref{sec:Mathieu} contains a collection of the formulae that
describe the motion of a single spinless particle in a periodic external
field. Here, we derive an exact expression for the one-particle propagator as well as
the modified L\"uscher zeta-function in the external field. Sect.~\ref{sec:proof}
is directly dedicated to the extraction of the local contribution to
the form factor. The proof of the Feynman-Hellman theorem for the
resonance form factor within the NREFT framework is also described here.
Finally, Sect.~\ref{sec:num} contains the results of the numerical study of the quantization
condition in an external field, which was carried out within a toy model. Note also that this
paper provides a proof of principle only. For this reason, we have simplified the physical problem
as much as possible. For example, we consider a non-relativistic case in detail, neglecting
relativistic corrections whatsoever in the beginning.
Moreover, to avoid clutter of indices, we restrict ourselves to the case of a single scalar
field and neglect all partial waves other than the S-wave.
All these effects can be taken into account in a rather straightforward
fashion, see a very brief discussion in Sect.~\ref{sec:relativistic}.

\section{Resonance form factor in the infinite volume}
\label{sec:infinite}

Let us consider a scalar non-relativistic particle with mass $m$, moving in
an external electromagnetic field $A^\mu(x)$. We shall further assume that
only $A^0(x)$ is different from zero,
and that it corresponds to the static field, i.e.,  $A^0(x)=A^0({\bf x})$.
The Lagrangian which
describes particles in this field consists of an infinite tower
of operators with increasing mass dimension that respect all symmetries, namely
rotational invariance, the discrete symmetries, and gauge invariance.
In the following, we shall
restrict ourselves to at most two particles in the initial and final states. Hence, the operators
in the Lagrangian should
contain at most two fields $\phi$ and two conjugated fields.
Furthermore, only terms up to first order in the coupling $e$ will be included in the
Lagrangian, since we are exclusively interested in the linear shift in the external field.

We shall start from the Lagrangian\footnote{For a review of the non-relativistic effective
  theories for hadrons see, e.g., Ref.~\cite{Gasser:2007zt}}
\eq\label{eq:Lagrangian}
   {\cal L}&=&\phi^\dagger\biggl(i\partial_t-m+eA^0+\frac{eC_R}{6m^2}\,\triangle A^0
   +\frac{\nabla^2}{2m}\biggr)\phi
   +C_0\phi^\dagger\phi^\dagger\phi\phi
   \nonumber\\[2mm]
   &+&C_2\biggl(\phi^\dagger\phi^\dagger(\phi\,{\gnabla}^2\!\!\phi)+\mbox{h.c.}\biggr)
   +\frac{e\kappa}{4}\,\phi^\dagger\phi^\dagger\phi\phi \triangle A^0\, ,
   \en
   where the Galilei-invariant derivative is defined as
   $a\gnabla b=\frac{1}{2}\,(a\nabla b-b\nabla a)$ and $\triangle$ denotes the Laplacian.
   Note that in the above Lagrangian we did not make an attempt to
   write down
   all possible terms up to a given order in the expansion in the inverse
   powers of $m$. Hence, the theory, defined
   by it, is only a model that nevertheless
   possesses all essential ingredients of the full
   theory. For the sake of clarity, we shall
   consider the proof on the basis of this model first, and address the
   general
   case very briefly only at the end\footnote{A brief comment about  gauge invariance is in order.
     The restrictions
     $A^0=A^0({\bf x})$ and ${\bf A}=0$ do not leave room for gauge transformations except a
     trivial shift of $A^0$ by a constant. In order to
   arrive at the Lagrangian given in Eq.~(\ref{eq:Lagrangian}), one has first
   to write down the most general gauge-invariant Lagrangian for arbitrary $A^\mu$, and choose a
   particular configuration of the external field afterwards. Note also that $\triangle A^0$ in
   Eq.~(\ref{eq:Lagrangian}) emerges from the
   gauge-invariant expression $-\nabla{\bf E}$, which reduces
   to $\triangle A^0$ for ${\bf A}=0$. }.

 The main aim of this section is to set up the framework for the evaluation of
 the resonance form factor in a theory described by the Lagrangian~(\ref{eq:Lagrangian}).
 The final result, given in Eq.~(\ref{eq:resFF}), can be derived in few steps.
 We start  from the two-particle scattering amplitude for the process
 $q_1+q_2\to p_1+p_2$ at $e=0$ (no external field).
 In the non-relativistic
 effective theory, this amplitude
 is given by a sum of bubble diagrams (we remind the reader that, for simplicity,
we focus on S-wave scattering only):
\eq\label{eq:T2}
T({\bf p},{\bf q};{\bf P};P^0)=\frac{8\pi}{m}\,
\biggl\{K(p,q)+K(p,q_0)\frac{iq_0}{1-iq_0K(q_0,q_0)}\,K(q_0,q)\biggr\},
\en
where  $p=|{\bf p}|$, $q=|{\bf q}|$, and
\eq\label{eq:K2}
q_0^2=m\biggl(P^0-2m-\frac{{\bf P}^2}{4m}\biggr)\, ,\quad\quad
K(p,q)=\frac{m}{8\pi}\,\biggl(4C_0-4C_2(p^2+q^2)\biggr)\, .
\en
Note that $P^0$ has an infinitesimal positive imaginary part $P^0\to P^0+i\varepsilon$ which,
for brevity, is never shown explicitly.
External particles are on mass shell: $p_i^0=m+{\bf p}_i^2/(2m)$
and $q_i^0=m+{\bf q}_i^2/(2m)$ for $i=1,2$.
Furthermore, the center-of-mass and relative momenta are given by
\eq\label{eq:CM}
{\bf P}={\bf p}_1+{\bf p}_2={\bf q}_1+{\bf q}_2\, ,\quad\quad
{\bf p}=\frac{{\bf p}_1-{\bf p}_2}{2}\, ,\quad\quad   
{\bf q}=\frac{{\bf q}_1-{\bf q}_2}{2}\, .
\en
On the energy shell, ${\bf p}^2={\bf q}^2=q_0^2$ and the total energy $P^0$ is given by
\eq
P^0=2m+\frac{{\bf P}^2}{4m}+\frac{q_0^2}{m}\, .
\en
The on-shell scattering amplitude takes the form 
\eq
\begin{displaystyle}
T(q_0)=\frac{8\pi/m}{K^{-1}(q_0,q_0)-iq_0}
=\frac{8\pi/m}{-1/a+rq_0^2/2+\cdots -iq_0}\, ,
\end{displaystyle}
\en
where
\eq
C_0=-\frac{2\pi a}{m}\, ,\quad\quad C_2=\frac{\pi ra^2}{2m}\, ,
\en
and $a,r$ denote the S-wave scattering length and effective range, respectively.

Let us now adjust the parameters $a,r$ so that there is a low-lying resonance
in the S-wave. In this case, the resonance pole position is determined from the equation:
\eq
-\frac{1}{a}+\frac{1}{2}\,rq_R^2-\sqrt{-q_R^2}=0\, .
\en
The choice of the minus sign in front of the square root corresponds to
the second Riemann sheet. 

Suppose that $a,r$ are chosen so that
the above equation has a solution with $\mbox{Re}\,q_R^2>0$,
$\mbox{Im}\,q_R^2<0$, with $|\mbox{Im}\,q_R^2|\ll
|\mbox{Re}\,q_R^2|\ll m^2$. This solution corresponds to a low-lying
resonance in the S wave.
In moving frames, the complex resonance energy is then given by
\eq
P^0_R=2m+\frac{{\bf P}^2}{4m}+\frac{q_R^2}{m}=\mbox{Re}\,P^0_R-\frac{i}{2}\,{\it\Gamma}_R\, ,
\en
where ${\it\Gamma}_R$ denotes the width of the resonance.

In the vicinity of the resonance pole, the two-body amplitude behaves as
\eq
T(q_0)&=&\frac{8\pi/m}{K^{-1}(q_0,q_0)-\sqrt{-q_0^2}}\to
\frac{Z}{q_0^2-q_R^2}+\mbox{regular terms}\, ,
\nonumber\\[2mm]
Z&=&\frac{8\pi/m}{\bigl[K^{-1}(q_R,q_R)\bigr]'-\bigl[\sqrt{-q_R^2}\bigr]'}\, ,
\en
where primes indicate derivatives with respect to the variable $q_0^2$, and $q_0^2=q_R^2$ is set at the end.
In the following, $Z$ will be referred to as the wave function renormalization constant of the
resonance. It is, in general, a complex quantity.

Let us now turn the coupling to the external field on, and consider the two-point function of a
particle in the external field up to $O(e)$:
\begin{align}
\begin{split}
S({\bf p},{\bf q};p^0)=&i\int dt\, d^3{\bf x}d^3{\bf y}\,
e^{ip^0t-i{\bf p}{\bf x}+i{\bf q}{\bf y}}
\langle 0|T\phi({\bf x},t)\phi^\dagger({\bf y},0)|0\rangle
\\
=&\,\frac{(2\pi)^3\delta^3({\bf p}-{\bf q})}{m+
  \dfrac{{\bf p}^2}{2m}-p^0}
+\frac{e{\it\Gamma}({\bf p},{\bf q})\tilde A^0({\bf p}-{\bf q})}
{\biggl(m+\dfrac{{\bf p}^2}{2m}-p^0\biggr)
  \biggl(m+\dfrac{{\bf q}^2}{2m}-p^0\biggr)}+O(e^2)\, ,
\end{split}
\end{align}
where
\eq
\tilde A^0({\bf p}-{\bf q})=\int d^3{\bf x} \,
e^{-i({\bf p}-{\bf q}){\bf x}}A^0({\bf x})
\en
is the Fourier-transform of the (static) scalar potential. Furthermore,
the one-particle form factor ${\it\Gamma}({\bf p},{\bf q};p^0)$ can be
directly read off from the Lagrangian,
\eq\label{eq:one_particle_vertex}
{\it\Gamma}({\bf p},{\bf q})=1-\frac{C_R}{6m^2}\,({\bf p}-{\bf q})^2\, ,
\en
where the quantity $C_R$ is related to the mean charge radius
through $C_R=m^2\langle r^2\rangle$. Note also that the
on-shell condition for the non-relativistic particles is
${\bf p}^2/(2m)={\bf q}^2/(2m)=p^0-m$.

Next, we turn to the definition of the resonance form factor.
This quantity can be defined through the expansion of the (equal-time)
four-point function in
the external field, similarly to the one-particle form factor obtained through
the expansion of the two-point function. This four-point function
is defined as
\eq\label{eq:fourpoint-external}
\tilde G({\bf p},{\bf P};{\bf q},{\bf Q};P^0)
&=&i\int dt\, d^3{\bf x}_1d^3{\bf x}_2d^3{\bf y}_1d^3{\bf y}_2\,
e^{iP^0t-i{\bf p}_1{\bf x}_1-i{\bf p}_2{\bf x}_2
  +i{\bf q}_1{\bf y}_1+i{\bf q}_2{\bf y}_2}
\nonumber\\[2mm]
&\times&
\langle 0|T\phi({\bf x}_1,t)\phi({\bf x}_2,t)
\phi^\dagger({\bf y}_1,0)\phi^\dagger({\bf y}_2,0)|0\rangle\, .
\en
In the absence of the external field, the equal-time four-point function
can be related to the two-particle scattering amplitude, considered above.
Writing $\tilde G=\tilde G_0+e\tilde G_1+O(e^2)$, we obtain
\begin{align}
\begin{split}
\tilde G_0({\bf p},{\bf P};{\bf q},{\bf Q};P^0)&=
\frac{(2\pi)^3\delta^3({\bf P}-{\bf Q})\,(2\pi)^3\bigl(\delta^3({\bf p}-{\bf q})
+\delta^3({\bf p}+{\bf q})\bigr)}
{2m+\dfrac{{\bf P}^2}{4m}+\dfrac{{\bf p}^2}{m}-P_0}
\\
&+\,\frac{(2\pi)^3\delta^3({\bf P}-{\bf Q})T({\bf p},{\bf q};{\bf P};P^0)}
{\biggl(2m+\dfrac{{\bf P}^2}{4m}+\dfrac{{\bf p}^2}{m}-P_0\biggr)
  \biggl(2m+\dfrac{{\bf P}^2}{4m}+\dfrac{{\bf q}^2}{m}-P^0\biggr)}\, ,
  \end{split}
\end{align}
where $T({\bf p},{\bf q};{\bf P};P^0)$ is the two-body amplitude,
introduced in Eqs.~(\ref{eq:T2}), (\ref{eq:K2}), and  ${\bf P}$, ${\bf Q}$
are the total three-momenta of the particle pairs in the initial and 
final states, respectively, see Eq.~(\ref{eq:CM}).
The four-point function has a simple pole at $P^0\to P^0_R$,
\eq\label{eq:fourpoint}
\tilde G_0({\bf p},{\bf P};{\bf q},{\bf Q};P^0)
\to(2\pi)^3\delta^3({\bf P}-{\bf Q})\frac{\Psi({\bf P},{\bf p})\bar\Psi({\bf Q},{\bf q})}{P^0-P_R^0}\, ,
\en
where
\eq
\Psi({\bf P},{\bf p})&=&
\frac{1}{2m+\dfrac{{\bf P}^2}{4m}+\dfrac{{\bf p}^2}{m}-P_0}\,
\sqrt{\frac{Z}{m}}\frac{K(p,q_R)}{K(q_R,q_R)} \, ,
\nonumber\\[2mm]
\bar\Psi({\bf Q},{\bf q})&=&\sqrt{\frac{Z}{m}}\frac{K(q,q_R)}{K(q_R,q_R)}\, 
\frac{1}{2m+\dfrac{{\bf Q}^2}{4m}+\dfrac{{\bf q}^2}{m}-P_0}\, .
\en
In analogy to the case of the two-particle bound states, we shall refer
to the quantity $\Psi$ as the ``wave function of a resonance.'' Note that
this does not have anything to do with the interpretation of a resonance
as a true quantum-mechanical state described by this ``wave
function'', but just represents a convenient brief name.

Next, it can be checked that, like true wave functions,  the quantities
$\Psi,\bar\Psi$ are normalized according to
\begin{align}\label{eq:norm}
\begin{split}
\frac{1}{2!}\,\int\frac{d^3{\bf p}}{(2\pi)^3}\,
\bar\Psi({\bf P},{\bf p})\Psi({\bf P},{\bf p})
  &=\frac{Z}{2mK^2(q_R,q_R)}\,
  \int\frac{d^3{\bf p}}{(2\pi)^3}\,\frac{K^2(p,q_R)}
  {\biggl(2m+\dfrac{{\bf P}^2}{4m}+\dfrac{{\bf p}^2}{m}-P_0\biggr)^2}
  \\
  &\hspace*{-2.cm}=\,-\frac{Zm}{8\pi K^2(q_R,q_R)}\,\frac{d}{dq_0^2}\biggl(
  K^2(q_0,q_R)\sqrt{-q_0^2}\biggr)\biggr|_{q_0^2=q_R^2}
  \\
  &\hspace*{-2.cm}=\,-\frac{Zm}{8\pi K^2(q_R,q_R)}\,\biggl(
  K'(q_R,q_R)+ K^2(q_R,q_R)\bigl[\sqrt{-q_R^2}\bigr]'\biggr)
  =1\, .
\end{split}
\end{align}
  Here, the factor $1/2!$ emerges from the Bose-symmetry.
  In the derivation, we have used the fact that the function $K(p,q)$ is
  real and symmetric (this, in its turn, stems from the hermiticity
  of the Hamiltonian) and, hence,
\begin{equation}
  \frac{d}{dq_0^2}\,K(q_0,q_R)\biggl|^{\phantom{\bigl|}}_{q_0^2=q_R^2\phantom{\biggl|}}
  =\frac{1}{2}\,\biggl[K(q_R,q_R)\biggr]'.
\end{equation}

  \begin{figure}[t]
    \begin{center}
      \includegraphics*[width=8.cm]{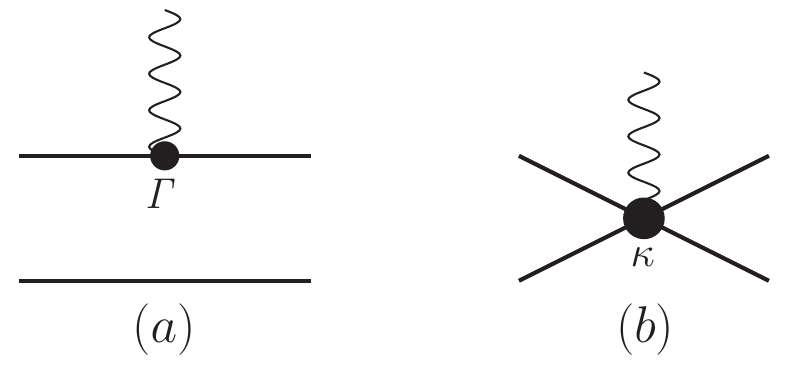}
      \caption{Diagrams contributing to the kernel $\tilde{\it\Gamma}$ which, convoluted with
        the wave functions $\bar\Psi,\Psi$, yields the resonance form factor, see
        Eq.~(\ref{eq:resFF}). The troublesome triangle diagram (a) in the finite-volume form factor,
        mentioned in the Introduction, emerges from the disconnected part, where the photon is attached
        to one of the particles, whereas the second one acts as a spectator. The fully connected
        diagram (b), where the photon is emanated from the four-particle vertex, is unproblematic.}
      \label{fig:Gammatilde}
    \end{center}
  \end{figure}

Up to order $e$, the equal-time Green function  takes the form
\eq
\tilde G_1({\bf p},{\bf P};{\bf q},{\bf Q};P^0)
&=&\frac{1}{(2!)^2}\int \frac{d^3{\bf p}'}{(2\pi)^3}\,\frac{d^3{\bf P}'}{(2\pi)^3}\,\frac{d^3{\bf q}'}{(2\pi)^3}\,\frac{d^3{\bf Q}'}{(2\pi)^3}\,
\tilde G_0({\bf p},{\bf P};{\bf p}',{\bf P}';P^0)
\nonumber\\[3mm]
&\times&
\tilde{\it\Gamma}({\bf p}',{\bf P}';{\bf q}',{\bf Q}')
\tilde G_0({\bf q}',{\bf Q}';{\bf q},{\bf Q};P^0)\, .
\en
At this order, the vertex $\tilde{\it\Gamma}$ is given by a sum of a finite number of diagrams,
shown in Fig.~\ref{fig:Gammatilde}:
\eq
\tilde{\it\Gamma}({\bf p},{\bf P};{\bf q},{\bf Q})
=\bar{\it\Gamma}({\bf p},{\bf P};{\bf q},{\bf Q})\tilde A^0({\bf P}-{\bf Q})\, ,
\en
where
\begin{align}
\begin{split}
&\bar{\it\Gamma}({\bf p},{\bf P};{\bf q},{\bf Q})=
-\kappa({\bf P}-{\bf Q})^2
\\
&+
\biggl\{(2\pi)^3\delta^3\biggl(\frac{({\bf P}-{\bf Q})}{2}+({\bf p}-{\bf q})\biggr)
{\it \Gamma}\biggl(\frac{{\bf P}}{2}-{\bf p},\frac{{\bf Q}}{2}-{\bf q}
\biggr)+
\left(\begin{array}{c}
{\bf p}\to-{\bf p}\cr{\bf q}\to-{\bf q}\cr
{\bf p}\to-{\bf p},\,{\bf q}\to-{\bf q}
\end{array}\right)\biggr\}\, ,
\label{eq:baritGamma}
\end{split}
\end{align}
and
$\it \Gamma$ is the one-particle vertex, given in
Eq.~(\ref{eq:one_particle_vertex}).

The quantity $\tilde G_1$ has a double pole in the variables $P^0,Q^0$
that is contained in the free Green functions $\tilde G_0$. The residue
at the double pole defines the form factor of the resonance:
\eq\label{eq:resFF}
F({\bf P},{\bf Q})=
\frac{1}{(2!)^2}\,\int \frac{d^3{\bf p}}{(2\pi)^3}\,
\frac{d^3{\bf q}}{(2\pi)^3}\,\bar\Psi({\bf P},{\bf p})
\bar{\it\Gamma}({\bf p},{\bf P};{\bf q},{\bf Q})
\Psi({\bf Q},{\bf q})\, ,
\en
where the energy is fixed at the resonance pole.
Using the normalization of the wave functions, given in Eq.~(\ref{eq:norm}), and the fact that,
due to the Bose-symmetry, these wave functions
are symmetric under ${\bf p}\to-{\bf p}$ and  ${\bf q}\to-{\bf q}$, respectively, it can
be immediately shown that the resonance form factor is
properly normalized at zero momentum transfer, as required by the Ward identity
(we remind the reader that the charge of the resonance is equal to $2e$). In Appendix~\ref{app:formfactor} we provide the explicit form of Eq.~(\ref{eq:resFF}) in dimensional regularization.

All parameters that are present in the Lagrangian~(\ref{eq:Lagrangian}) enter
the expression~(\ref{eq:resFF}) as well. Namely, the wave functions $\Psi,\bar\Psi$
contain the elastic two-particle scattering parameters $C_0,C_2$, whereas
the kernel $\bar{\it\Gamma}$ depends on the parameter $C_R$ that describes the
single particle form factor, as well as the coupling $\kappa$, characterizing the contact
term. There will be more couplings, if higher-order derivative terms, higher partial waves, etc.,
are included, but the general pattern is already clear. All these couplings should be determined on the lattice, on the same configurations. In order to determine the elastic scattering phase, related to $C_0,C_2$, one may use standard L\"uscher approach for the two-body scattering at $e=0$. The value of $C_R$ can be established by calculating the single particle form factor by using either the standard method or the Feynman-Hellmann theorem. At the order
we are working, only a single constant $\kappa$ remains unknown. Below
it will be shown, how this constant can be fixed in the external field.

The framework that we considered in this Section is not new and represents a properly
adapted version of the Mandelstam formalism~\cite{Mandelstam,HuangWeldon}, which is used
to define form factors of {\em stable} particles. The purpose of such a detailed treatment was
to set the stage for a similar calculation in a finite volume.  In the following,
it will be demonstrated that using the Feynman-Hellmann
theorem in a periodic external field, one arrives exactly at the quantity
defined by Eq.~(\ref{eq:resFF}) in the infinite-volume limit.

\section{Single particle in a periodic
  external field}
\label{sec:Mathieu}

\subsection{Solutions of the Mathieu equation}

Up to this point, the discussion was carried out for a generic static external
field $A^0({\bf x})$. In order to inject a momentum between the initial and final states on the
lattice, it is convenient to consider a spatially periodic field
\eq
A^0({\bf x})=A_0\cos(\boldsymbol{\omega}{\bf x})\, ,\quad\quad
\boldsymbol{\omega}=(0,0,\omega)\, .
\en
Here, for convenience, we have chosen the vector $\boldsymbol{\omega}$ in the direction of
the $z$-axis. Furthermore, we
project all vectors onto the direction of $\boldsymbol{\omega}$:
for instance, the position vector has the components ${\bf x}=({\bf x}_\perp,x_\parallel)$, where
${\bf x}_\perp$, $x_\parallel$ denote the components perpendicular and parallel to the $z$-axis,
respectively.

In this section, we shall derive a closed expression of the two-point function of
the field $\phi$ in the external field.
We are working here in a cubic box with a spatial elongation $L$ (the
time elongation is assumed to be infinite). Periodic boundary conditions
are imposed in the spatial directions. As a result, the three-momenta
of the particles as well as the frequency $\omega$ are quantized:
\eq
&&{\bf p}=\frac{2\pi}{L}\,{\bf n}\, ,\quad {\bf n}\in\mathbb{Z}^3\, , \,\,
\text{ and } \quad
\omega=\frac{2\pi}{L}\,N\, ,\quad N\in\mathbb{Z}\, .
\en
Let us denote by $|1\rangle$ a state with a single particle in the periodic
field. The matrix element of the field operator between the vacuum and
the one-particle state defines the Schr\"odinger wave function
\eq
\Phi({\bf x},t)=\langle 0|\phi({\bf x},t)|1\rangle\, .
\en
The wave function obeys a differential equation that can be obtained
by using the equations of motion for the field $\phi({\bf x},t)$:
\eq\label{eq:differential}
\biggl(i\partial_t+e{\it\Gamma}A_0\cos(\omega x_\parallel)-m
+\frac{\nabla^2}{2m}\biggr)\Phi({\bf x},t)=0\, .
\en
Here,
\eq
{\it\Gamma}={\it\Gamma}(\boldsymbol{\omega})
=1-\frac{C_R}{6m^2}\,\omega^2
\en
is the single-particle form factor evaluated at the three-momentum transfer
$\omega$. Note that, after factorizing  Eq.~(\ref{eq:differential})
by using an ansatz  $\Phi({\bf x},t)=e^{-iEt+i{\bf p}_\perp{\bf x}_\perp}f(x_\parallel)$,
this equation can be reduced to a so-called Mathieu equation for the function $f(x_\parallel)$.
The (unnormalized) solutions of  Eq.~(\ref{eq:differential}) that obey
periodic boundary conditions are given by 
\eq\label{eq:solutions}
\Phi({\bf x},t)&=&e^{-iEt+i{\bf p}_\perp{\bf x}_\perp}{\rm me}_{\nu_i+2n}(z,q)\, ,
\nonumber\\[2mm]
z&=&\frac{\omega x_\parallel}{2}\, ,\quad\quad
q=-\frac{4me{\it\Gamma}A_0}{\omega^2}\, .
\en
where ${\rm me}_{\nu_i+2n}(z,q)$ denotes the
Mathieu function and the index $\nu_i+2n$, $n\in \mathbb{Z}$, $i=1,\ldots,N$ labels
the eigenfunctions of the Mathieu differential equation corresponding to the eigenvalues
$\lambda_{\nu_i+2n}(q)$~\cite{McLachlan,NIST}. Details are given in  App.~\ref{app:Mathieu}. 

The completeness condition for the solutions of the Mathieu equation takes
the form:
    \eq
    \frac{1}{\pi}\sum_{i=1}^N\sum_{n=-\infty}^\infty
  {\rm me}_{\nu_i+2n}(z,q){\rm me}_{\nu_i+2n}(-z',q)=
  N\sum_{k=-\infty}^\infty 
  \delta(z-z'-\pi k N)\, ,
  \en
  with the $\nu_i$ as given in Eq.~(\ref{eq:nui}).

The propagator of the particle $\phi$ in the external field is defined as:  
\eq
S({\bf x},{\bf y};E)=i\int_{-\infty}^{+\infty} dt e^{iEt}
\langle 0|T\phi({\bf x},t)\phi^\dagger({\bf y},0)|0\rangle\, ,
\en
and it is given by the sum over the eigenfunctions (spectral representation):
\begin{align}
\begin{split}
\label{eq:SA_x}
S({\bf x},{\bf y};E)&=\frac{1}{L^3}\,
\sum_{{\bf p}_\perp}\sum_{i=1}^N\sum_{n=-\infty}^\infty
\frac{e^{i{\bf p}_\perp({\bf x}_\perp-{\bf y}_\perp)}}
{m+\dfrac{{\bf p}_\perp^2}{2m}+\dfrac{\omega^2}{8m}\,
  \lambda_{\nu_i+2n}(q)-E}
\\
&\times
{\rm me}_{\nu_i+2n}\biggl(\frac{\omega x_\parallel}{2},q\biggr)
{\rm me}_{\nu_i+2n}\biggl(-\frac{\omega y_\parallel}{2},q\biggr)\, .
\end{split}
\end{align}
Indeed, it can be directly verified that
\eq
\biggl(E+eA_0\cos(\omega x_\parallel)-m+\frac{\nabla^2}{2m}\biggr)
S({\bf x},{\bf y};E)
=-\sum_{{\bf m}\in\mathbb{Z}^3}
\delta^3({\bf x}-{\bf y}-{\bf m}L)\, .
\en
The propagator can be expanded in powers of $e$ (this corresponds
to the Taylor expansion in the parameter $q$). Up to $O(e)$, the
result takes the expected simple form:
\eq\label{eq:perturbative}
S({\bf p},{\bf q};E)&=&\int^Ld^3{\bf x}d^3{\bf y}
e^{-i{\bf p}{\bf x}
  +i{\bf q}{\bf y}}S({\bf x},{\bf y};E)
\nonumber\\[2mm]
&=&L^3\left\{\frac{\delta^3_{{\bf p}{\bf q}}}{m+\dfrac{{\bf p}^2}{2m}-E}+\frac{1}{2}\,eA_0{\it\Gamma}
\frac{\delta^3_{{\bf p}-\boldsymbol{\omega},{\bf q}}
  +\delta^3_{{\bf p}+\boldsymbol{\omega},{\bf q}}}
{\biggl(m+\dfrac{{\bf p}^2}{2m}-E\biggr)
  \biggl(m+\dfrac{{\bf q}^2}{2m}-E\biggr)}\right\}+O(e^2)\, .
\nonumber\\
\en
The proof of this equation is given in Appendix~\ref{app:expansion}.

\subsection{The energy shift in the periodic field}
\label{sec:energyshift-1}

The spectrum of a particle in an external periodic field is determined by
the poles of the propagator. 
Performing the Fourier transform using Eq.~(\ref{eq:Fourier}),
the propagator can be rewritten in the following form:
\eq
S({\bf p},{\bf q};E)&=&L^3\delta^2_{{\bf p}_\perp,{\bf q}_\perp}
\sum_{i=1}^N\sum_{n=-\infty}^\infty\sum_{a,b=-\infty}^\infty
C_{2a}^{\nu_i+2n}(q)C_{2b}^{\nu_i+2n}(q)
\nonumber\\[2mm]
&\times&\frac{\delta_{-p_\parallel,\frac{\omega}{2}\,(\nu_i+2n+2a)}
\delta_{-q_\parallel,\frac{\omega}{2}\,(\nu_i+2n+2b)}}
{m+\dfrac{{\bf q}_\perp^2}{2m}+\dfrac{\omega^2}{8m}\,
  \lambda_{\nu_i+2n}(q)-E}\, .
\en
Here, the coefficients $C_{2a,2b}^{\nu_i+2n}(q)$ ($a,b\in \mathbb{Z}$) are the same as in
Eq.~(\ref{eq:Fourier}), and their explicit form does not matter here.
We see now that, instead of one pole, the propagator in the external field
has a tower of poles. This was expected, because the periodic external
field carries the momentum $\boldsymbol{\omega}$. Consequently,
the three-momentum is not conserved in such a field, and ${\bf q}={\bf p}+\ell\boldsymbol{\omega}$,
where $\ell\in\mathbb{Z}$ is an integer.
In addition, since the particle interacts
with the field, the energies (pole positions) are slightly displaced from the
non-interacting values corresponding to $\lambda_{\nu_i+2n}(0)=(\nu_i+2n)^2$
and are determined through the equation
\eq
E=m+\dfrac{{\bf p}_\perp^2}{2m}+\frac{\omega^2}{8m}\,
\lambda_{\nu_i+2n}(q)\, .
\en
The crucial point is that $\lambda_{\nu_i+2n}(q)=(\nu_i+2n)^2+O(q^2)$ for all values
of $\nu_i+2n$ except $(\nu_i+2n)=\pm 1$. In this case,
\begin{equation}\label{eq:lambda1-1}
\lambda_1(q)=1+q+O(q^2)\, ,\quad\quad
\lambda_{-1}(q)=1-q+O(q^2)\, .
\end{equation}
In the following, for simplicity, we shall take
${\bf p}_\perp={\bf q}_\perp=0$ and determine the {\em lowest}
eigenvalue in the sectors with different values of $N$.
(Note that the integer number $N$ characterizes the momentum transfer
in the external field vertex, in units of ${2\pi}/{L}$.)
The components $p_\parallel,q_\parallel$ are given by
$p_\parallel=2\pi n_\parallel/L$ and $q_\parallel=2\pi n'_\parallel/L$.
In the sectors
with $N=1,2$ the argument goes as follows:
\subsubsection*{\underline{$N=1$:}}
In this case, $\nu_i=0$, and we have
  \eq
  -n_\parallel=n+a\, ,\quad\quad 
  -n'_\parallel=n+b\, .
  \en
  For any choice of $n_\parallel,n'_\parallel$, we may find $a,b$ so that $n=0$. Hence, the lowest
  eigenvalue is $\lambda_{\nu_i+2n}(q)=\lambda_0(q)=O(q^2)$.

\subsubsection*{\underline{$N=2$:}}
  In this case, we have $\nu_i=0,1$ and
  \eq
   -n_\parallel=\nu_i+2n+2a\, ,\quad\quad
   -n'_\parallel=\nu_i+2n+2b\, .
   \en
   This means that $n_\parallel,n'_\parallel$ should be either odd
   or even. If both are chosen to be even, then $\nu_i=0$ should be fulfilled
   and $n=0$ is allowed. Then, the lowest eigenvalue is $\lambda_0(q)$.
   On the other hand, if $n_\parallel,n'_\parallel$ are odd, then $\nu_i=1$
   is picked up. Since $n$ is integer, $\nu_i+2n$ is odd and $\lambda_0(q)$
   never appears. The lowest eigenvalues are then $\lambda_{\pm 1}(q)$.
   Assuming $A_0>0$ and $q>0$, we get $\lambda_1(q)<\lambda_1(-q)$
   and the lowest energy level will be at
\eq
E=m+\frac{\omega^2 {\lambda_1(q)}}{8m}=m+{\frac{\omega^2}{8m}}-\frac{1}{2}\,eA_0{\it\Gamma}\, .
\en
Hence, differentiating the pole shift with respect to $e$, one gets the
particle form factor
${\it\Gamma}={\it\Gamma}(\boldsymbol{\omega})$.
This is exactly the case considered in Ref.~\cite{QCDSF:2017ssq}:
one places the charged particle in the periodic external field with
$\omega=4\pi/L$, and considers the Breit frame
${\bf p}=-{\bf q}=-\boldsymbol{\omega}/2$. Then, the linear derivative
of the shift of the lowest energy level with respect to the coupling to
the external field
yields the form factor at the momentum transfer 
$\boldsymbol{\omega}$. Hence, our result confirms and extends
the findings of Ref.~\cite{QCDSF:2017ssq} to different incoming and outgoing momenta,
as well as to the higher values of $N$.

\section{Two-particle scattering in the periodic external field}
\label{sec:proof}

\subsection{L\"uscher equation}

\begin{figure}[th!]
  \begin{center}
    \includegraphics*[width=11.cm]{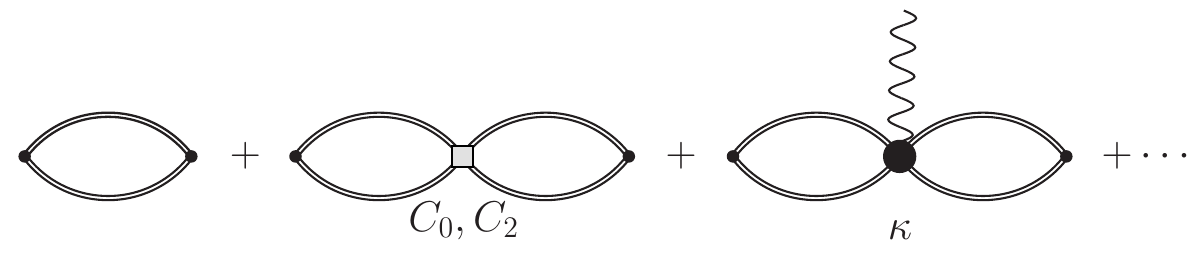}
    \caption{Two-point function of the composite field $\phi^2(x)$. Double
      lines denote the full one-particle propagator in the external field, see
      Fig.~\ref{fig:composite_prop-1}.
      The diagrams, in which the external field is attached to the four-particle
      vertex, are explicitly included.}
    \label{fig:composite_prop}
  \end{center}
\end{figure}

In this section, we shall derive the counterpart of the L\"uscher equation
in the external field, which allows one to extract the contact coupling,
$\kappa$, from the finite-volume energy spectrum.
To this end, let us consider the two-point function of the composite fields $\phi^2,[\phi^\dagger]^2$:
\eq
D({\bf P},{\bf Q};t)=i\int^Ld^3{\bf x}d^3{\bf y}
e^{-i{\bf P}{\bf x}+i{\bf Q}{\bf y}}
\langle 0|T\phi^2({\bf x},t)[\phi^\dagger({\bf y},0)]^2|0\rangle\, .
\en
The diagrams that contribute to this quantity are shown in Fig.~\ref{fig:composite_prop}.
These are reminiscent of the diagrams in the absence of an external field, with two differences:
a) the particle propagators in these diagrams are the full ones
that include the summation of all external field insertions
in these propagators, and b) in addition to the conventional four-particle vertices,
there are vertices with the external field attached (the pertinent
operator comes with the coupling $\kappa$ in the Lagrangian). Below, we shall study the
implications of these modifications.

\begin{figure}[th!]
  \begin{center}
    \includegraphics*[width=11.cm]{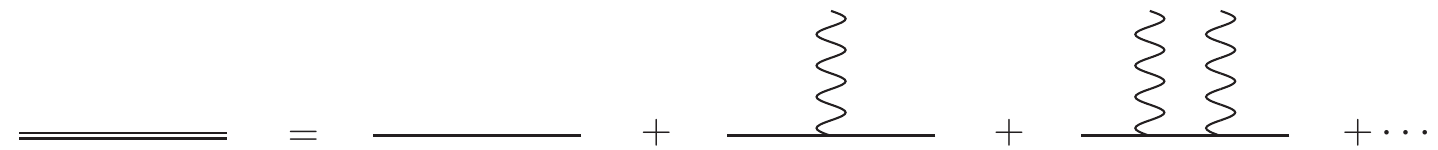}
    \caption{Full one-particle propagator in the external field.}
    \label{fig:composite_prop-1}
  \end{center}
\end{figure}

First, note that, since the three-momentum is not conserved
in the presence of the external field, the two-point function is no more diagonal in the
incoming/outgoing total three-momenta ${\bf Q}$ and ${\bf P}$. Instead of
an overall factor $L^3\delta^3_{{\bf P}{\bf Q}}$ it contains a tower of terms
with $L^3\delta^3_{{\bf P}+\ell\boldsymbol{\omega},{\bf Q}}$ and $\ell\in\mathbb{Z}$, $\ell \neq 0$.
However, since each momentum flip proceeds through the interaction
with the external field and thus adds one power of the coupling $e$,
the terms that multiply  $L^3\delta^3_{{\bf P}+\ell\boldsymbol{\omega},{\bf Q}}$
start at $O(e^\ell)$. Hence, at a given order in $e$, the quantity
$D({\bf P},{\bf Q};E)$ is a band matrix with the indices ${\bf P},{\bf Q}$.
Note also that here we do not attempt {\it a priori} to expand the particle propagator
in powers of $e$, since such an expansion cannot be easily justified on the real axis
of the energy.

Let us consider the elementary loop diagram in Fig.~\ref{fig:composite_prop}.
If both four-particle vertices in such a diagram do not contain derivatives
of the field $\phi$, such a loop is given merely by a convolution of two propagators
in the external field
\eq\label{eq:luescherzeta}
   {\it\Pi}({\bf P},{\bf Q};E)=\int\frac{dp^0}{2\pi i}\,
   \int^Ld^3{\bf x}d^3{\bf y}\,e^{-i{\bf P}{\bf x}+i{\bf Q}{\bf y}}
   S({\bf x},{\bf y};p^0)S({\bf x},{\bf y};E-p^0)\, .
\en
Here, $S({\bf x},{\bf y};E)$ is defined by Eq.~(\ref{eq:SA_x}).

The situation is slightly more complicated in case of  vertices
with derivatives, e.g., the vertex that is proportional to the coupling $C_2$
in Eq.~(\ref{eq:Lagrangian}). In the case with no external field and
using dimensional regularization, it is possible to ``pull out'' the
derivatives acting on the internal lines and transform them into the external
momenta. The difference is an off-shell term, which cancels the denominator
in the loop, leaving a low-energy polynomial that vanishes after integration
in dimensional regularization. The above fact allows one to derive the L\"uscher equation
in a very simple manner.\footnote{The final result is the same in all regularizations but the
  use of  dimensional regularization
makes the derivation particularly simple.} When the external field
is switched on, ``pulling out'' the derivatives leads to an extra term
that depends on the external field. Indeed, using Eq.~(\ref{eq:luescherzeta})
together with Eq.~(\ref{eq:Mathieu2}), one can easily show that
\begin{align}
\begin{split}
\label{eq:Pi-deriv}
&\int^Ld^3{\bf x}d^3{\bf y}\,e^{-i{\bf P}{\bf x}+i{\bf Q}{\bf y}}
\biggl\{S({\bf x},{\bf y};p^0)(\stackrel{\rightarrow}{\nabla}_x
-\stackrel{\leftarrow}{\nabla}_x)^2S({\bf x},{\bf y};E-p^0)\biggr\}
\\
&=\int^Ld^3{\bf x}d^3{\bf y}\,e^{-i{\bf P}{\bf x}+i{\bf Q}{\bf y}}
\\
&\times
\frac{1}{L^3}\,
\sum_{{\bf p}_\perp}\sum_{i=1}^N\sum_{n=-\infty}^\infty
\frac{e^{i{\bf p}_\perp({\bf x}_\perp-{\bf y}_\perp)}
  {\rm me}_{\nu_i+2n}\biggl(\dfrac{\omega x_\parallel}{2},q\biggr)
{\rm me}_{\nu_i+2n}\biggl(-\dfrac{\omega y_\parallel}{2},q\biggr)}
     {m+\dfrac{{\bf p}_\perp^2}{2m}+\dfrac{\omega^2}{8m}\,
       \lambda_{\nu_i+2n}(q)-p^0-i\varepsilon}
\\
&\times\biggl(-2{\bf p}_\perp^2-2{\bf q}_\perp^2+{\bf P}^2-\frac{\omega^2}{2}\,(\lambda_{\nu_i+2n}(q)+\lambda_{\nu_j+2m}(q))-8me{\it\Gamma}A^0({\bf x})\biggr)
\\
&\times
\frac{1}{L^3}\,
\sum_{{\bf q}_\perp}\sum_{j=1}^N\sum_{m=-\infty}^\infty
\frac{e^{i{\bf q}_\perp({\bf x}_\perp-{\bf y}_\perp)}
  {\rm me}_{\nu_j+2m}\biggl(\dfrac{\omega x_\parallel}{2},q\biggr)
{\rm me}_{\nu_j+2m}\biggl(-\dfrac{\omega y_\parallel}{2},q\biggr)}
     {m+\dfrac{{\bf q}_\perp^2}{2m}+\dfrac{\omega^2}{8m}\,
       \lambda_{\nu_j+2m}(q)-E+p^0-i\varepsilon}\, .
\end{split}
\end{align}
The expression in the brackets, which is present in the numerator,
can be rewritten as
     \eq
   &-&2{\bf p}_\perp^2-2{\bf q}_\perp^2+{\bf P}^2-\frac{\omega^2}{2}\,(\lambda_{\nu_i+2n}(q)+\lambda_{\nu_j+2m}(q))-8me{\it\Gamma}A^0({\bf x})  
\nonumber\\[2mm]
&=&-4m\biggl(m+\frac{{\bf p}_\perp^2}{2m}+\frac{\omega^2}{8m}\,\lambda_{\nu_i+2n}(q)-p^0\biggr)
-4m\biggl(m+\frac{{\bf q}_\perp^2}{2m}+\frac{\omega^2}{8m}\,\lambda_{\nu_j+2m}(q)
-E+p^0\biggr)
\nonumber\\[2mm]
&-&4m\biggl(E-2m-\frac{{\bf P}^2}{4m}\biggr)-8me{\it\Gamma}A^0({\bf x})\, .
\en
The first two terms cancel with one of the denominators in Eq.~(\ref{eq:Pi-deriv}).
Using dimensional regularization, it can be argued that these two terms give a
vanishing contribution to the integral. The third term corresponds to ``pulling
out'' the derivatives on the internal lines. Only the last term is new
and shows that, in case of a non-vanishing external field, there is an
additional contribution. Physically, this corresponds to the four-particle vertex
with the external field attached (topologically equivalent to the one that contains the
coupling $\kappa$). In other words,
pulling out the derivatives is equivalent to adjusting the coefficients
of such terms\footnote{Note, however, that this additional term does not carry
the momenta ${\bf P},{\bf Q}$ and, in particular, does not vanish at
$({\bf P}-{\bf Q})^2=0$. This indicates that pulling all derivatives out of the loop,
checking Ward identities as well as the normalization of the
form factor at zero momentum transfer can become technically
complicated, albeit gauge invariance still holds.}.

Carrying out this procedure consistently in all loops, it is seen that
the full two-point function $D$ obeys the equation
\eq\label{eq:luescher-external}
\frac{1}{4}\,D_{{\bf P}{\bf Q}}(E)=\frac{1}{2}\,{\it \Pi}_{{\bf P}{\bf Q}}(E)
+\frac{1}{L^6}\sum_{{\bf P}'{\bf Q}'}\frac{1}{2}\,{\it \Pi}_{{\bf P}{\bf P}'}(E)
X_{{\bf P}'{\bf Q}'}(E)\frac{1}{4}\,D_{{\bf Q}'{\bf Q}}(E)\, .
\en
Here, for convenience, we have used matrix notation, considering the momenta
${\bf P},{\bf Q},\ldots$ as the indices. The kernel $X$ is given by
\eq\label{eq:kernelX}
X_{{\bf P}{\bf Q}}(E)=L^3\delta^3_{{\bf P}{\bf Q}}X^{(0)}_{\bf P}(E)
    +\frac{e}{2}\,L^3(\delta^3_{{\bf P}+\boldsymbol{\omega},{\bf Q}}
    +\delta^3_{{\bf P}-\boldsymbol{\omega},{\bf Q}})
    X^{(1)}_{{\bf P}{\bf Q}}(E)
+O(e^2)\, ,
      \en
      where
\eq\label{eq:X0X1}      
X^{(0)}_{\bf P}(E)&=&4C_0-8mC_2\biggl(E-2m-\frac{{\bf P}^2}{4m}\biggr)+\cdots\, ,
  \nonumber\\[2mm]
  X^{(1)}_{{\bf P}{\bf Q}}(E)&=&-\kappa\omega^2A_0-16mC_2{\it\Gamma}A_0+\cdots\, .
    \en
    Note that the second term in $X^{(1)}$ emerges after pulling out the
    derivatives.

    The derivation of the L\"uscher equation is now straightforward. The energy
    levels are determined by the equation
    \eq\label{eq:luescher_ext}
    \det {\cal M}=0\, ,\quad\quad
         {\cal M}_{{\bf P}{\bf Q}}(E)=[X_{{\bf P}{\bf Q}}(E)]^{-1}
         -\frac{1}{2}\,{\it\Pi}_{{\bf P}{\bf Q}}(E)\, .
         \en
Furthermore, up to first order in $e$, the inverse of the matrix $X$ is given by
\eq\label{eq:Xm1}
[X_{{\bf P}{\bf Q}}(E)]^{-1}&=&L^3\delta^3_{{\bf P}{\bf Q}}k({\bf P};E)
    -\frac{e}{2}\,L^3(\delta^3_{{\bf P}+\boldsymbol{\omega},{\bf Q}}
    +\delta^3_{{\bf P}-\boldsymbol{\omega},{\bf Q}})
    k({\bf P};E)X^{(1)}_{{\bf P}{\bf Q}}(E)k({\bf Q};E)
\nonumber\\[2mm]
    &+&O(e^2)\, ,
\en
where
\eq
k({\bf P};E)&=&\biggl(4C_0-8mC_2\biggl(E-\frac{{\bf P}^2}{4m}\biggr)+\cdots\biggr)^{-1}
=\frac{m}{8\pi}\,\biggl(-\frac{1}{a}+\frac{1}{2}\,rq_0^2({\bf P};E)+\cdots\biggr)\, ,
\nonumber\\[2mm]
q_0^2({\bf P};E)&=&m\biggl(E-2m-\frac{{\bf P}^2}{4m}\biggr)\, .
\en
As seen from the above equation, at leading order in $e$, the inverse of the kernel
reduces to the well-known expression $q_0\cot\delta(q_0)$. This was of course
expected from the beginning. The $O(e)$ corrections to the kernel can be
calculated perturbatively in a consistent manner. At this order, they are
characterized by a single unknown effective coupling $\kappa$.

Furthermore, if ${\bf P}=-{\bf Q}$ (Breit frame), the non-diagonal term
in Eq.~(\ref{eq:Xm1}) can be rewritten as
\eq\label{eq:kappa-r}
k({\bf P};E)X^{(1)}_{{\bf P}{\bf Q}}(E)k(-{\bf P};E)
&=&-k^2({\bf P};E)\kappa\omega^2A_0-\frac{m^2}{4\pi}\,{\it\Gamma}A_0\frac{dK^{-1}(q_0,q_0)}{dq_0^2}+\cdots
\nonumber\\[2mm]
&=&-k^2({\bf P};E)\kappa\omega^2A_0-\frac{m^2}{8\pi}\,{\it\Gamma}A_0r+\cdots\, .
\en
In other words, at this order, everything is expressed in terms of the effective-range
parameters $a,r$ and the coupling $\kappa$.

Equation~(\ref{eq:luescher_ext}) is one of our main results, namely the L\"uscher equation
in the presence of an external field. In contrast to the conventional
L\"uscher equation, which reduces to a single equation in the absence of
partial-wave mixing, Eq.~(\ref{eq:luescher_ext}) results from
the matrix equation that connects sectors with different momenta ${\bf P},{\bf Q}$.
This happens because three-momentum is not a conserved quantity in the case considered here. 

In order to make the equations tractable, a truncation should be applied. Let us consider
the Breit frame again, with
${\bf P}=-{\bf Q}=\boldsymbol{\omega}/2$. If $e=0$, ${\cal M}$ is a diagonal matrix,
whose matrix elements at ${\bf P}={\bf Q}=\pm \boldsymbol{\omega}/2$ linearly
vanish at the energies that correspond to the finite-volume spectrum of a system in a
frame moving with a momentum ${\bf P}$. Turning the external field on, it is seen that the energy
levels split and continuously shift from these values\footnote{In fact, as we shall see
  later, the structure of the spectrum at $e\neq 0$ is more complicated. There  exist
  ``fake'' states which do not have a counterpart at $e=0$.}. It can be straightforwardly
checked that, at order $e$,
it suffices to consider a $2\times 2$ matrix with ${\bf P}=\pm\boldsymbol{\omega}/2$ and 
${\bf Q}=\pm\boldsymbol{\omega}/2$. Adding more rows and columns to this matrix
shifts the spectrum in higher orders only.

It is important to realize that the only missing piece in our knowledge of the resonance
form factor is the contact contribution, which is proportional to the constant $\kappa$ in
our example. Everything else is known: the form factor in the impulse approximation
(this corresponds to the triangle diagram in Fig.~\ref{fig:triangle}a) is determined
through the known form factors of individual particles. This way, the coupling $\kappa$ can
be extracted by fitting the data to the energy spectrum obtained from
the L\"uscher equation in the external field, Eq.~(\ref{eq:luescher_ext}). Unlike
measured matrix elements of the external current, $\kappa$, by definition,
may contain only exponentially suppressed contributions in a finite volume. Hence,
this method allows one to circumvent the problem of the irregular $L$-dependence,
mentioned in the Introduction\footnote{This statement should be clarified by an example. Suppose that
one calculates the finite-volume energy spectrum in an ``exact'' theory (be this QCD
or relativistic EFT), and then extracts $\kappa$ from this spectrum by using the
NREFT setting described in this paper. The extracted quantity $\kappa=\kappa(L)$ will
depend on $L$. We state that the difference $\kappa(L)-\kappa(\infty)=O(e^{-\mu L})$
(modulo a prefactor that may contain powers of $L$), where $\mu$ is some scale
given by a multiple of the lightest mass in the system (here, the only available
scale is the particle mass $m$ itself). In this context, one might term this statement,
which applies to all effective couplings in NREFT, as the finite-volume counterpart
of the Appelquist-Carazzone decoupling theorem.}.

\subsection{The L\"uscher zeta-function in the external field; perturbative
  expansion}

We shall now provide an explicit expression for the loop function ${\it\Pi}$, defined
in Eq.~(\ref{eq:luescherzeta}). 
Carrying out the integration over the transverse momenta and the energy, we get
\eq
{\it\Pi}({\bf P},{\bf Q};E)=L^2\delta^2_{{\bf P}_\perp{\bf Q}_\perp}
\bar{\it\Pi}(P_\parallel,Q_\parallel;{\bf P}_\perp;E)\, ,
\en
where
\eq\label{eq:Luescher-modified}
&&\bar{\it\Pi}(P_\parallel,Q_\parallel;{\bf P}_\perp;E)=\frac{1}{L^4}\,\sum_{{\bf p}_\perp}
\sum_{i,j=1}^N\sum_{n,m=-\infty}^\infty\int_0^L dx_\parallel \int_0^L dy_\parallel\,
D_{in,jm}({\bf p}_\perp;{\bf P}_\perp;E)e^{-iP_\parallel x_\parallel+iQ_\parallel y_\parallel}
\nonumber\\[2mm]
&&\times\,{\rm me}_{\nu_i+2n}\biggl(\frac{\omega x_\parallel}{2},q\biggr)
{\rm me}_{\nu_i+2n}\biggl(-\frac{\omega y_\parallel}{2},q\biggr)
{\rm me}_{\nu_j+2m}\biggl(\frac{\omega x_\parallel}{2},q\biggr)
{\rm me}_{\nu_j+2m}\biggl(-\frac{\omega y_\parallel}{2},q\biggr)\, ,\quad\quad
\en
and
\eq
D_{in,jm}({\bf p}_\perp;{\bf P}_\perp;E)
=\frac{1}{2m+\dfrac{{\bf p}_\perp^2}{2m}+\dfrac{({\bf P}-{\bf p})_\perp^2}{2m}
  +\dfrac{\omega^2}{8m}\,(\lambda_{\nu_i+2n}(q)+ \lambda_{\nu_j+2m}(q))-E}\, .\quad\quad
\en
Below, we shall consider the perturbative expansion of this expression
in powers of the coupling $e$ (or, equivalently, the quantity $q$). The
reason for this is twofold. First, in the standard method of evaluating
the resonance form factor, the matrix element between the two-particle
scattering states is calculated on the lattice. This corresponds to taking
exactly $O(e)$ term in the perturbative expansion. Hence, expanding the result
in $e$, one may establish a closer relation between the ``standard'' approach
and the approach which is proposed in the present paper. The second reason
is practical. The full expression of the L\"uscher function in the external
field is quite cumbersome and is not well suited for numerical evaluation. The
expansion allows one to arrive at a much simpler expression. One should
be however aware of pitfalls, see below.

In what follows, we shall display the result of the calculation of this quantity at
first order in the parameter $q$. Note that the energy denominator
depends on $q$ as well since
$\lambda_{\pm 1}(q)=1\pm q+O(q^2)$. Hence, the perturbative expansion fails
at the energies where the pertinent denominators vanish at $O(q^0)$. For
this reason, along with the ``perturbative'' result, we also present the
``exact'' one, obtained by expanding the numerator in powers of $q$ but
keeping the $O(q)$ terms in the denominator unexpanded. The implications
of using the ``perturbative'' result instead of the ``exact'' one are
also considered in detail.

\begin{figure}[t]
  \begin{center}
    \includegraphics*[width=0.45\linewidth]{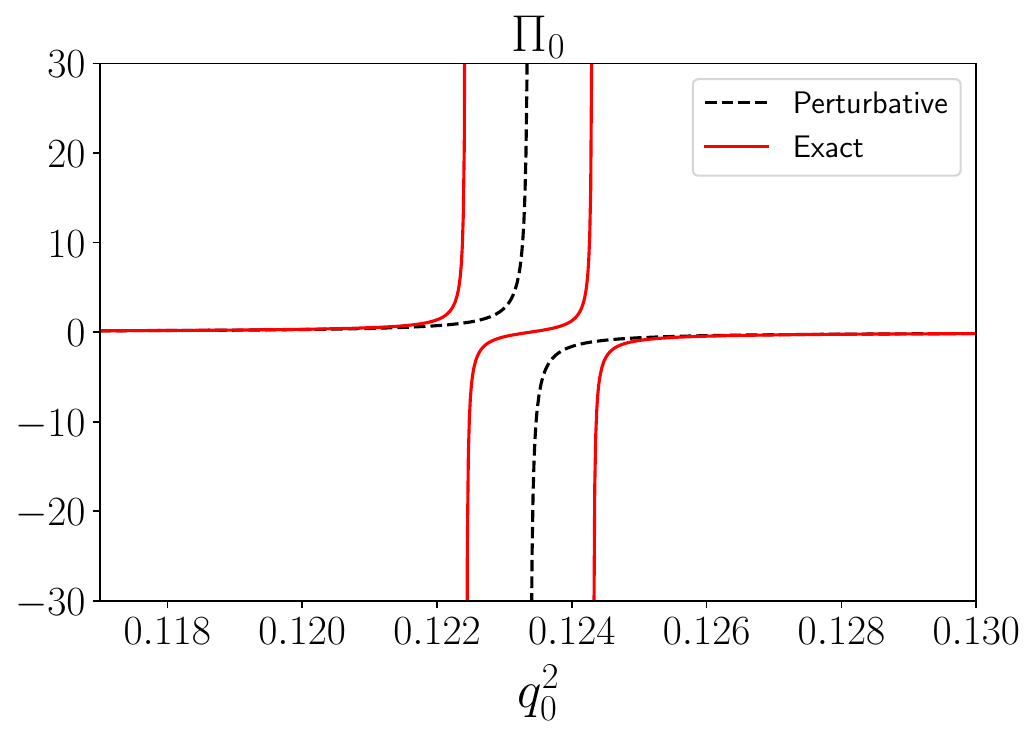}
    \includegraphics*[width=0.45\linewidth]{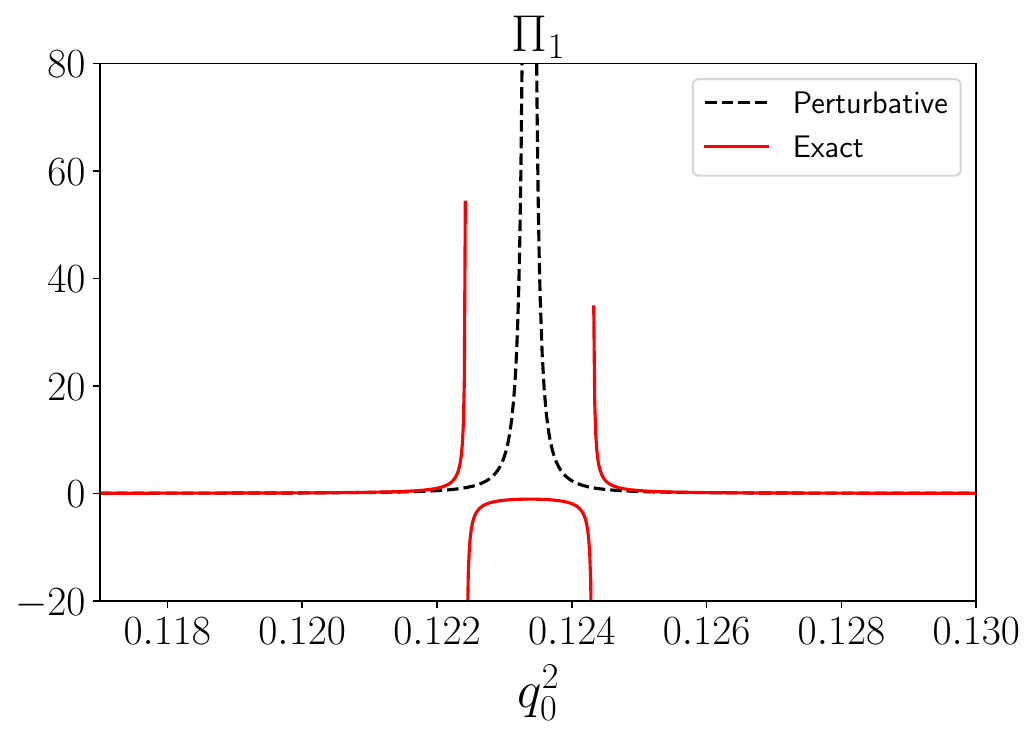}
    \caption{The diagonal and non-diagonal elements of the matrix ${\it\Pi}$, given
      in Eqs.~(\ref{eq:Piprime}) and (\ref{eq:Pi2prime}). The quantity $q_0^2$ is given
      in Eq.~(\ref{eq:K2}). It is seen that the quantity ${\it\Pi}_1^{(1)}$ (``perturbative'') develops
      a double pole instead of the two separated simple poles found in ${\it\Pi}_1$ (``exact'').
      Since this figure serves illustrative purpose only, we do not
      specify the values of the parameters in calculations. The units on the vertical axis are
      arbitrary.}
    \label{fig:Pi}
  \end{center}
  \end{figure}

The initial and final momenta ${\bf P},{\bf Q}$ in the above equations are
arbitrary. Below, we shall use the notation $P_\parallel=a\omega/2$,
$Q_\parallel=b\omega/2$. For simplicity, we shall further
restrict ourselves to the $2\times 2$ matrix
with $a,b=\pm 1$ (recall that ${\bf P}_\perp={\bf Q}_\perp$).
The detailed derivation can be found in Appendix~\ref{app:Luescher-modified}.
The pertinent elements of the matrix ${\it \Pi}_{ab}$ are
denoted as $\tilde{\it\Pi}_{11}=\tilde{\it\Pi}_{-1,-1}\doteq{\it \Pi}_0$ and
$\tilde{\it\Pi}_{1,-1}=\tilde{\it\Pi}_{-1,1}\doteq{\it \Pi}_1$.
The ``exact'' and ``perturbative'' results are denoted
by $\Pi_{0,1}$ and $\Pi'_{0,1}$, respectively: 
\eq\label{eq:Piini}
{\it\Pi}_0={\it\Pi}^{(1)}_0+{\it\Pi}^{(2)}_0\, ,\quad\quad
{\it\Pi}_1={\it\Pi}^{(1)}_1+{\it\Pi}^{(2)}_1\, ,
\en
where
\begin{align}
\begin{split}
\label{eq:Piprime}
{\it\Pi}^{(1)}_0&=\frac{1}{L^3}\,\sum_{\bf p}\frac{1}{2m+\dfrac{{\bf p}^2}{2m}
  +\dfrac{({\bf P}-{\bf p})^2}{2m}-E}\, ,
\\
{\it\Pi}^{(1)}_1&=-\frac{\omega^2q}{4m}\,\frac{1}{L^3}\,
\sum_{\bf p}\frac{1}{\left(2m+\dfrac{{\bf p}^2}{2m}
  +\dfrac{({\bf P}-{\bf p})^2}{2m}-E\right)\left(2m+\dfrac{{\bf p}^2}{2m}
  +\dfrac{({\bf Q}-{\bf p})^2}{2m}-E\right)}\, ,\quad\quad
\end{split}
\end{align}
and
\begin{align}
\begin{split}
\label{eq:Pi2prime}
{\it\Pi}^{(2)}_0&=\frac{1}{L^3}\,\sum_{{\bf p}_\perp}\left\{
\frac{1+\dfrac{q}{4}}{\dfrac{\omega^2}{8m}\,(1+q)+\dfrac{\omega^2}{2m}-W}
+\frac{1-\dfrac{q}{4}}{\dfrac{\omega^2}{8m}\,(1+q)-W}\right.
\\
&+\left.\frac{1-\dfrac{q}{4}}{\dfrac{\omega^2}{8m}\,(1-q)+\dfrac{\omega^2}{2m}-W}
+\frac{1+\dfrac{q}{4}}{\dfrac{\omega^2}{8m}\,(1-q)-W}
-\frac{2}{\dfrac{\omega^2}{8m}+\dfrac{\omega^2}{2m}-W}
-\frac{2}{\dfrac{\omega^2}{8m}-W}\right\}\, ,
\\
{\it\Pi}^{(2)}_1&=\frac{1}{L^3}\,\sum_{{\bf p}_\perp}\left\{
\frac{\dfrac{\omega^2q}{4m}}{\left(\dfrac{\omega^2}{8m}-W\right)^2}
+\left(
  \frac{1}{\dfrac{\omega^2}{8m}\,(1+q)-W}
  -\frac{1}{\dfrac{\omega^2}{8m}\,(1-q)-W}
\right)\right.
\\
&+
  \frac{q}{4}\,\left(
\frac{1}{\dfrac{\omega^2}{8m}\,(1+q)+\dfrac{\omega^2}{2m}-W}
-\frac{2}{\dfrac{\omega^2}{8m}\,(1+q)-W}\right.
\\
&+
\left.\left.
    \phantom{\frac{\dfrac{\omega^2q}{4m}}{\left(\dfrac{\omega^2}{8m}-W\right)^2}}
    \hspace*{-2.4cm}
    \frac{1}{\dfrac{\omega^2}{8m}\,(1-q)+\dfrac{\omega^2}{2m}-W}
-\frac{2}{\dfrac{\omega^2}{8m}\,(1-q)-W}
-\frac{2}{\dfrac{\omega^2}{8m}+\dfrac{\omega^2}{2m}-W}
+\frac{4}{\dfrac{\omega^2}{8m}-W}\right)\right\}\, .
\end{split}
\end{align}
Note also that the following notation is used:
\eq
W=E-2m-\frac{{\bf p}_\perp^2}{2m}-\frac{({\bf P}-{\bf p})_\perp^2}{2m}\, .
\en
Some comments are in order now.
As already said, the quantities ${\it\Pi}^{(1)}_{0,1}$ can be obtained straightforwardly
by using the perturbative expansion of the one-particle propagator,
see Eq.~(\ref{eq:perturbative}). Namely,
${\it\Pi}^{(1)}_0$ leads to the L\"uscher zeta-function, and ${\it\Pi}^{(1)}_1$
is nothing but the triangle diagram (or, the so-called $G$-function, in the
approach of Refs.~\cite{Briceno:2015tza,Baroni:2018iau}). Hence, the relation to
the ``standard'' approach is clearly visible.
However, we already know that
this expansion fails in the vicinity of the free particle poles. Indeed, instead of one simple
pole at $q=0$, the one-particle propagator possesses two poles at $q\neq 0$, which are located
symmetrically on both sides. Expanding the denominator in powers of $q$, one gets one
double pole instead of two single poles,
separated by a distance $2q$. This is schematically shown in Fig.~\ref{fig:Pi}.
where the names ``exact'' and ``perturbative'' refer to
${\it\Pi}_{0,1}$ and ${\it\Pi}^{(1)}_{0,1}$, respectively. Furthermore, it is worth noting that, formally,
${\it\Pi}^{(2)}_{0,1}$ are at least of order $q^2$ and can be neglected. We have seen, however that such
an approximation does not suffice in the vicinity of the singularities. Another observation is that,
in the infinite-volume limit, which can be performed for energies below the two-particle threshold,
the quantities  ${\it\Pi}^{(2)}_{0,1}$ behave like $L^{-1}$ for large $L$ modulo exponential corrections.

\subsection{``Exact'' vs. ``perturbative'' solution}

  A very interesting question arises, namely, whether the solutions of the L\"uscher
  equation in the external field are the same up to terms of order $e^2$, if
  one replaces ${\it\Pi}_{0,1}$ by ${\it\Pi}^{(1)}_{0,1}$ (we remind the reader that the difference
  between these quantities is {\em formally} of order $e^2$). The answer to this question is positive, and will be discussed below.

Let us take, for simplicity, ${\bf P}_\perp=0$ and try
  to obtain a solution in the vicinity of $E=2m+\omega^2/(8m)$.
  The  quantities ${\it\Pi}_{0,1}$ exhibit here the following behavior
  \eq
  {\it\Pi}_{0,1}=\frac{c_{0,1}^+(q)}{2m+\dfrac{\omega^2}{8m}\,(1+q)-E}
  +\frac{c_{0,1}^-(q)}{2m+\dfrac{\omega^2}{8m}\,(1-q)-E}+\bar{\it\Pi}_{0,1}\, ,
  \en
  where $\bar{\it\Pi}_{0,1}$ is a smooth function of $E$   in the vicinity of $E=2m+\omega^2/(8m)$
  and the coefficients
  \eq
  c_0^\pm=\frac{1}{L^3}\,\biggl(1\mp\frac{q}{4}\biggr)\, ,\quad\quad
  c_1^\pm=\pm\frac{1}{L^3}\,\biggl(1\mp\frac{q}{2}\biggr)\, ,
\en
can be directly
read off from Eqs.~(\ref{eq:Piprime}) and (\ref{eq:Pi2prime}).
Thus, the L\"uscher equation in this case can be reduced to two algebraic
  equations of the type
  \eq\label{eq:Taylor}
  \mathscr{G}_\pm^{-1}(E)\doteq\frac{c_+(q)}{2m+\dfrac{\omega^2}{8m}\,(1+q)-E}
  +\frac{c_-(q)}{2m+\dfrac{\omega^2}{8m}\,(1-q)-E}+f_\pm(E,q)=0\, .
  \en
  Here, $c_\pm=-\frac{1}{2}\,(c^\pm_0\pm c^\pm_1)$ and
  $f_\pm(E,q)$ are some smooth functions of their arguments which contains
  $\bar{\it\Pi}_{0,1}$ as well as the elements of the matrix $X$, Eq.~(\ref{eq:X0X1}).
  Hence, one could expand
  $f_\pm(E,q)$ in Taylor series in $E$ in the vicinity of the unperturbed level
  and solve the obtained equations iteratively. At lowest order, $f_\pm(E,q)$ are
  just constants
  and one gets quadratic equations for $E$ which can be easily solved. Indeed, 
rewriting the above equation as
 \eq\label{eq:G-1}
  \mathscr{G}_\pm^{-1}(E)&=&
  \frac{mc_+}{q_+^2-q_0^2}
  +\frac{mc_-}{q_-^2-q_0^2}+f_\pm=0\, ,
  \nonumber\\[2mm]
  q_\pm^2&=&\frac{\omega^2}{16}\,(1\pm 2q)\, ,
  \en
  one gets two roots
  \eq
  q_0^2&=&q_1^2+\frac{m(c_+ + c_-)}{2f_\pm}
  \pm\frac{1}{2f_\pm}\,\sqrt{D}\, ,
\nonumber\\[2mm]
  q_1^2&=&\frac{1}{2}\,(q_+^2+q_-^2)=\frac{\omega^2}{16}\, ,
\nonumber\\[2mm]
  D&=&\left(2f_\pm q_1^2+m(c_++c_-)\right)^2
    -4f_\pm(f_\pm q_+^2q_-^2+mc_+q_-^2+mc_-q_+^2)\, .  
  \en
  Expanding this solution in powers of $q$ one gets a pair of solutions
  that differ by the choice of sign in front of the square root
  \eq\label{eq:q02-exp}
  q_0^2=q_1^2+\frac{m(c_++c_-)}{2f_\pm}\pm
  \frac{m(c_++c_-)}{2f_\pm}\,\biggl(1-\frac{f_\pm(q_-^2-q_+^2)(c_+-c_-)}
  {m(c_++c_-)^2}\biggr)+\cdots\, .
  \en
  Choosing the ``$+$'' sign,
  one may verify that one gets exactly
  the same result as first expanding   Eq.~(\ref{eq:Taylor}) in powers of $q$ and then
  solving it with respect to $E$. Higher orders in the expansion in $E$ can be treated in the
  similar fashion. On the other hand, with the choice of the ``$-$'' sign,
  one arrives at the solution $q_0^2=q_1^2+O(q)$. As we shall see below, this
  corresponds to an unphysical solution.

  Albeit the argument, given above, proves that the linear dependence of the energy
  levels is not altered by using perturbative expansion in $\Pi_{0,1}$, the situation for
  a finite $e$ is not that clear. It cannot be excluded that the structure
  of the energy levels is qualitatively different, until $e$ becomes sufficiently small
  (later, we shall demonstrate this explicitly).
  For this reason, using the
  exact solution for $\Pi_{0,1}$ in the fit is preferable. 

  \subsection{Residua}

  As mentioned earlier, the perturbative expansion contains pitfalls.
  Here we shall consider one of these. Namely, it will be shown that
  the use of the
  {\em expanded} L\"uscher equation might lead to the ``fake'' poles. The
  residua of these poles are however of order $e^2$ and could be thus neglected
  at the order one is working. In the actual fit to the lattice data, one should
  carefully identify such poles and exclude them from the analysis.

Below, for simplicity, we restrict ourselves
  to the case of the $2\times 2$ matrix already
  considered above. The quantity $\mathscr{G}(E)$, defined
  in Eq.~(\ref{eq:Taylor}), contains single poles at $q_0^2=q_n^2$, which are given,
  in particular, by Eq.~(\ref{eq:q02-exp}).  
  In the vicinity of such a pole,
  \eq
  \mathscr{G}(E)=\frac{\mathscr{Z}_n}{q_n^2-q_0^2}+\mbox{regular}\, .
  \en
  Differentiating both sides of the above equation with respect to $q_0^2$, one may easily ensure that
  the residuum at the pole, $\mathscr{Z}_n$, is given by the derivative of
  $\mathscr{G}_\pm^{-1}(E)$ at the pole:
\eq\label{eq:Zn-1}
  \mathscr{Z}_n^{-1} =-\lim_{q_0^2\to q_n^2} [\mathscr{G}_\pm^{-1}(E)]'\, .
\en
  On the other hand, from Eq.~(\ref{eq:G-1})
  one gets
  \eq\label{eq:Gpm-1}
   [\mathscr{G}_\pm^{-1}(E)]'=
  \frac{mc_+}{(q_+^2-q_0^2)^2}
  +\frac{mc_-}{(q_-^2-q_0^2)^2}+f'_\pm\, .
  \en
  Equation~(\ref{eq:q02-exp}) describes two solutions, which differ by the choice
  of the sign in front of the last term. Taking into account the fact that
  $c_+\pm c_-=O(1)$ and $q_\pm^2-q_1^2=O(q)$, one immediately gets that
  the quantity  $\mathscr{Z}$ is of order 1 and order $q^2$ for the first and the
  second solution, respectively. This demonstrates that the second solution
  is an
  artifact of the approximations used, since the terms of order $q^2$ in the numerators
  have been systematically neglected.

\subsection{Extracting the resonance pole}
\label{sec:extracting}

In the infinite volume momenta are no more quantized. However,
the conservation of the three-momentum, ${\bf P}={\bf Q}\pm\ell\boldsymbol{\omega}$, implies
that the two-point function of the composite fields
for a fixed $\boldsymbol{\omega}$
still obeys the matrix equation~(\ref{eq:luescher-external}). In this equation, however,
${\bf P},{\bf Q},\boldsymbol{\omega}$ are no more
restricted to  integer multiples of $2\pi/L$. Furthermore, the kernel $X$
is the same (modulo replacing the Kronecker deltas by the Dirac delta-functions), and the
loop function ${\it\Pi}$ is replaced by its infinite-volume counterpart that amounts to replacing
the sum over the loop momenta by an integral.

A crucial point is that we can use the perturbative
expansion~(\ref{eq:perturbative}) in the coupling $e$. The reason is that, in order to find
the position of the resonance pole,
we are going to solve the infinite-volume analog of Eq.~(\ref{eq:luescher-external}) in the
complex plane, where the energy denominator, appearing in the loop, is not singular
and the perturbative expansion is justified. However, as seen above, when
solving the L\"uscher equation on the real axis, the perturbative
series diverges in the vicinity of the singularities.  Hence, in a finite volume,
it is preferable to work with the full expression of the L\"uscher zeta-function in the external field\footnote{One should stress here once more that one
is forced to exclusively use perturbative expressions within the ``standard''
approach. From the discussion above it is however clear that both approaches
are algebraically equivalent at $O(e)$ (as it should be).}.

Up to $O(e)$ terms, the loop function ${\it\Pi}$ in the infinite
volume takes the form:
\begin{align}
\begin{split} \label{eq:loopPi}
{\it\Pi}({\bf P},{\bf Q};E)&=(2\pi)^3\delta^3({\bf P}-{\bf Q})
{\it\Pi}_0({\bf P};E)
\\
&+eA_0{\it\Gamma}\,(2\pi)^3\bigl[
\delta^3({\bf P}+\boldsymbol{\omega}-{\bf Q})
+\delta^3({\bf P}+\boldsymbol{\omega}-{\bf Q})\bigr]
{\it\Pi}_1({\bf P},{\bf Q};E)+O(e^2)\, ,\quad\quad\quad
\end{split}
\end{align}
where
\eq
{\it\Pi}_0({\bf P};E)&=&\int\frac{dp^0}{2\pi i}\int\frac{d^3{\bf p}}{(2\pi)^3}\,\frac{1}{\biggl(m+\dfrac{{\bf p}^2}{2m}-p^0\biggr)
  \biggl(m+\dfrac{({\bf P}-{\bf p})^2}{2m}-E+p^0\biggr)}
\nonumber\\[2mm]
&=&\frac{m}{4\pi}\,\biggl[-m\biggl(E-2m-\frac{{\bf P}^2}{4m}\biggr)\biggr]^{1/2}\, ,
\nonumber\\[2mm]
{\it\Pi}_1({\bf P},{\bf Q};E)&=&\int\frac{dp^0}{2\pi i}\int\frac{d^3{\bf p}}{(2\pi)^3}\,\frac{1}{\biggl(m+\dfrac{{\bf p}^2}{2m}-p^0\biggr)
  \biggl(m+\dfrac{({\bf P}-{\bf p})^2}{2m}-E+p^0\biggr)}
\nonumber\\[2mm]
&\times&
\frac{1}{\biggl(m+\dfrac{({\bf Q}-{\bf p})^2}{2m}-E+p^0\biggr)}
\nonumber\\[2mm]
&=&-\frac{m^2}{8\pi}\,
\int_0^1dx\,\frac{1}{\sqrt{m(2m+{\bf P}^2/(4m)-E)+\frac{1}{4}\,\omega^2x(1-x)}}
\nonumber\\[2mm]
&=&-\frac{m^2}{2\pi\omega}\,\arcsin\frac{\omega}{\sqrt{16m(2m+{\bf P}^2/(4m)-E)+\omega^2}} \label{eq:piinfvol}
\, .
\en
Note that the sign convention in front of the square roots in the
above expressions corresponds to the choice of the second Riemann
sheet.

The procedure for determining the position of the pole on the second
Riemann sheet is as follows. First, one uses the finite-volume energy
levels in the Breit frame, ${\bf P}=-{\bf Q}=\pm\boldsymbol{\omega}/2$, to extract
the parameters of the Lagrangian using the L\"uscher equation with an external field,
Eq.~(\ref{eq:luescher-external}). In our case, there is a single unknown parameter $\kappa$.
Next, solving the same equation in the infinite volume, using Eq.~(\ref{eq:piinfvol}),
with the extracted values of the couplings, one determines the position of the pole on the second sheet.
In this manner, one could study the dependence of the pole position with $e$.
It can be seen that a pair of poles emerges which move in opposite
directions as $e$ increases. At first order of $e$, they move with the same rate.

\subsection{Relation to the resonance form factor}
\label{sec:relation}

From the previous discussion,
the infinite-volume two-point function in the external field possesses
poles on the second Riemann sheet. In the vicinity of a pole,
the residue factorizes. In the Breit frame, one has:
\eq
D({\bf P},{\bf Q};E)\to(2\pi)^3\delta^3({\bf P}-{\bf Q}\pm\boldsymbol{\omega})
\frac{\Phi({\bf P})\bar\Phi(-{\bf P})}{P^0({\bf P})-P_R^0({\bf P},e)}\, .
\en
Here, we have explicitly indicated the dependence on the parameter $e$.

Next, we shall differentiate both sides of the above equation with respect
to $e$ and set $e=0$ at the end
(because we are interested only in the terms linear in $e$).
The most singular term (a double pole) comes
from differentiating the denominator. Hence,
\eq\label{eq:derivative1}
\frac{d}{de}\,D({\bf P},{\bf Q};E)\biggr|_{e=0}&\to&
(2\pi)^3\delta^3({\bf P}-{\bf Q}\pm\boldsymbol{\omega})
\frac{\Phi({\bf P})\bar\Phi(-{\bf P})}{(P^0({\bf P})-P_R^0({\bf P},0))^2}\,
\frac{dP_R^0({\bf P},e)}{de}\biggr|_{e=0}
\nonumber\\[2mm]
&+&\mbox{less singular terms.}
\en
On the other hand, the quantity $D$ can be identically
rewritten as $D=DD^{-1}D$. Differentiating with respect to $e$, one gets
$\dfrac{d}{de}\,D=-D\biggl(\dfrac{d}{de}\,D^{-1}\biggr)D$.
The quantity $D$ at $e=0$ has a pole
\eq
D({\bf P},{\bf Q};E)\to(2\pi)^3\delta^3({\bf P}-{\bf Q})
\frac{\Phi({\bf P})\bar\Phi({\bf P})}{P^0({\bf P})-P_R^0({\bf P},0)}\, .
\en
Taking into account the definition~(\ref{eq:fourpoint-external}) and
comparing with Eq.~(\ref{eq:fourpoint}), one can straightforwardly read
off the relation between the quantity $\Phi({\bf P})$ and the wave
function $\Psi({\bf P},{\bf p})$, introduced in Sect.~\ref{sec:infinite}:
\eq\label{eq:PsiPhi}
\Phi({\bf P})=\int\frac{d^3{\bf p}}{(2\pi)^3}\,\Psi({\bf P},{\bf p})\, .
\en
An important remark is in order. In the limit $e=0$, the three-momentum
is conserved and, hence, one can establish the relation between $\Psi$ and
$\Phi$ only for ${\bf P}={\bf Q}$. However, the fact that the residue at
the pole factorizes, enables one to write down the residue for
${\bf P}\neq {\bf Q}$ as well. In simple cases like the one considered here, the factorization at the pole
can be verified explicitly, carrying out the truncation
in the ${\bf P},{\bf Q}$ space and inverting the resulting matrix.

Finally, using $D^{-1}=\frac{1}{2}\,{\it\Pi}^{-1}-\frac{1}{4}\,X$ and
taking into account
Eqs.~(\ref{eq:kernelX}) and (\ref{eq:loopPi}), one gets
\eq\label{eq:derivative2}
&&\frac{d}{de}\,D({\bf P},{\bf Q};E)\to
-(2\pi)^3\delta^3({\bf P}-{\bf Q}\pm\boldsymbol{\omega})
\frac{\Phi({\bf P})\bar\Phi({\bf P})\,\Phi(-{\bf P})\bar\Phi(-{\bf P})}{(P^0({\bf P})-P_R^0({\bf P},0))^2}\,
\nonumber\\[2mm]
& \times&
\bigl[-\frac{1}{2}\,A_0{\it\Pi}_0^{-1}({\bf P},E_R)
{\it\Gamma}{\it\Pi}_1({\bf P},{\bf -P};E_R){\it\Pi}_0^{-1}(-{\bf P};E_R)
-\frac{1}{8}\,X^{(1)}_{{\bf P},-{\bf P}}(E_R)\bigr]\, .
\en
Comparing Eqs.~(\ref{eq:derivative1}) and (\ref{eq:derivative2}),
we finally obtain:
\eq\label{eq:derivative3}
&&\frac{dP_R^0({\bf P})}{de}\biggr|_{e=0}
\nonumber\\[2mm]
&=&\frac{1}{2}\,\bar\Phi({\bf P})\bigl[{\it\Pi}_0^{-1}({\bf P},E_R)
A_0{\it\Gamma}{\it\Pi}_1({\bf P},{\bf -P};E_R){\it\Pi}_0^{-1}(-{\bf P};E_R)
+\frac{1}{4}\,X^{(1)}_{{\bf P},-{\bf P}}(E_R)\bigr]\Phi(-{\bf P})\, .\quad\quad
\en
In order to prove that this expression is the same
as Eq.~(\ref{eq:resFF}), let us first assume that $C_2=0$ and
use ${\tilde A^0({\bf P}-{\bf Q})=\frac{1}{2}\,(2\pi)^3(\delta^3({\bf P}-{\bf Q}
  +\boldsymbol{\omega})+\delta^3({\bf P}-{\bf Q}-\boldsymbol{\omega})}$. When $C_2=0$, the
integration over the relative momenta in Eq.~(\ref{eq:resFF}) is performed trivially,
yielding Eq.~(\ref{eq:derivative3}) (note that Eq.~(\ref{eq:PsiPhi})
should be used to prove this relation). If $C_2\neq 0$, in analogy
to what was done before, one has to pull
out the derivatives acting on the internal lines. Then, the expression
for $X^{(1)}$ will be modified and one arrives again at
Eq.~(\ref{eq:X0X1}).  Equations~(\ref{eq:resFF}) and (\ref{eq:derivative3})
are also equivalent in this case. Finally, we arrive at our final result that looks remarkably simple:
\eq\label{eq:main}
\frac{1}{2}\,A_0F({\bf P},-{\bf P})=\frac{dP_R^0({\bf P})}{de}\biggr|_{e=0}\, .
\en
In other words, in the Breit frame the resonance form factor is given
by the derivative of the resonance pole position with respect to the coupling constant
with the external field\footnote{As already mentioned, each pole $e=0$ splits into
two, moving in the opposite direction at equal speed, when the external field is turned on.
Choosing another pole yields just a different sign in Eq.~(\ref{eq:main}).}.

To summarize, all what is needed to compute the resonance form factor is the contact
contribution (at the lowest order, this is parameterized by a single coupling constant,
$\kappa$). The latter can be determined by fitting directly the energy levels in the
external field\footnote{For instance, it could be advantageous to fit the quantity
  $\Delta\doteq \langle E\rangle_{\phi^2}-2\langle E\rangle_\phi$, calculated on the lattice
  in  the presence of the external field. This quantity describes the energy shift of the two-particle
  state caused by the interactions between them and might be more sensitive to the small effects coming
  from contact interactions parameterized by $\kappa$.}. The resonance form factor can be then
calculated using
Eqs.~(\ref{eq:baritGamma}) and (\ref{eq:resFF}). Hence, extracting the resonance
pole first and using then a Feynman-Hellmann theorem is even superfluous.
However, the direct analogy with the Feynman-Hellmann theorem for the
form factors of stable particles is still remarkable. 

Extracting the contact contribution could, however, be complicated, since this contribution
contains suppression factors. For example, from Eq.~(\ref{eq:kappa-r}) it is seen that the
contribution containing $\kappa$ is multiplied by a factor $k^2({\bf P};E)$. In the case
of a shallow and narrow resonance, this approximately equals to $q_0^2$. In addition,
owing to gauge invariance, a factor $\omega^2$ is present. This, however, is not an obstacle
for the extraction of the form factor, since the same suppression factors also emerge in the
expression of the latter in the infinite volume. In other words, if the quantity $\kappa$,
determined on the lattice, is zero within the error bars, this simply means that the form factor
at this accuracy is given only by the impulse approximation.

\subsection{Relativistic corrections, higher partial waves and all that}
\label{sec:relativistic}

In this section we briefly consider the generalization of the above approach to
higher orders in the momentum expansion. This is needed, in particular,
to render the approach applicable to the study of the problems where relativistic effects cannot
be neglected. The inclusion of the higher-derivative interaction terms (an analog of the term with
$C_2$), which also describe higher partial waves, as well as derivative four-particle interaction
with the external field (similar to the coupling $\kappa$), proceeds relatively straightforwardly
and will not be considered here. A single non-trivial piece is the modification of the Lagrangian
in the single particle sector. As it is known, derivative insertions in the non-relativistic
propagators should be summed up to all orders, in order to arrive at a correct dispersion relation.
We shall try to do the same in presence of the external field below.

In general, writing down all terms in the one-particle sector is a complicated task (in higher
orders)  and can be carried out order by order in the expansion in the inverse mass.
Matching should be performed in the same setting, order by order in the expansion. The
situation simplifies dramatically, if we additionally restrict ourselves to terms that are
linear in $e$. These should be matched to the relativistic form factor $F_\mu(p',p)=ie(p'_\mu+p_\mu)F(t)$,
with $t=(p'-p)^2$. In this case, the form of the Lagrangian can be read off directly
from the matching condition and takes the form (the differential operators act on
everything right to them):
\eq{\cal L}&=&\phi^\dagger\biggl(i\partial_t-W+e{\it\Gamma}
\frac{1}{\sqrt{2W}}\,(WA^0({\bf x})+A^0({\bf x})W)\frac{1}{\sqrt{2W}}\biggr)\phi
\nonumber\\[2mm]
&+&\mbox{terms with four fields}\, .
\en
Here, $W=\sqrt{m^2-\triangle}$ denotes the relativistic energy operator,
and ${\it\Gamma}(\boldsymbol{\omega})=F(-\omega^2)$.
The equation for the one-particle wave function takes the form
  \eq
  \biggl(i\partial_t-W+e{\it\Gamma}
  \frac{1}{\sqrt{2W}}\,(WA^0({\bf x})+A^0({\bf x})W)\frac{1}{\sqrt{2W}}\biggr)\Phi({\bf x},t)=0\, .
  \en
  Using Eq.~(\ref{eq:separation}), this equation can be rewritten as
  \eq\label{eq:Mathieu-rel}
  \biggl(E-W_\perp+e{\it\Gamma}\frac{1}{\sqrt{2W_\perp}}\,(W_\perp A^0({\bf x})+A^0({\bf x})W_\perp)
  \frac{1}{\sqrt{2W_\perp}}\biggr)\bar\Phi(z)=0\, ,
  \en
  where
  \eq
  W_\perp=\sqrt{m^2+{\bf p}_\perp^2+\frac{4}{\omega^2}\,\frac{d^2}{dz^2}}\, .
  \en  
  Albeit Eq.~(\ref{eq:Mathieu-rel}) does not have
  the form of the Mathieu equation,
  at  first order in $e$ it can be reduced to it through the redefinition
  of the wave function:
  \eq\label{eq:transform}
  \bar\Phi(z)=\sqrt{2W_\perp}\biggl(1-\frac{e{\it\Gamma}}{\sqrt{2W_\perp}}\,A^0({\bf x})\biggr) \bar\Phi'(z)\, .
  \en
  The equation for the transformed wave function can then be rewritten as:
  \eq
  \biggl(E-W_\perp+e{\it\Gamma}A^0({\bf x})+O(e^2)\biggr)\bar\Phi'(z)=0\, ,
  \en
  or similarly,
    \eq
  &&  \biggl((E+e{\it\Gamma}A^0({\bf x}))^2-W_\perp^2+O(e^2)\biggr)\bar\Phi'(z)
\nonumber\\[2mm]
  &=&
  \biggl(E^2-m^2-{\bf p}_\perp^2+\frac{4}{\omega^2}\,\frac{d^2}{dz^2}+\frac{2Ee{\it\Gamma}}{\omega}\,\cos 2z+O(e^2)\biggr)\Phi'(z)
    =0\, .
  \en
  This is an equation of the Mathieu type, where the non-relativistic
  dispersion law (as in Eq.~(\ref{eq:Mathieu1})) is replaced by the relativistic expression
  $E^2-m^2+{\bf p}_\perp^2$. Note, however, that the parameter $q$ in this equation depends on
  the eigenvalue $E$, so the solutions can be found only numerically with an iterative procedure.
  Once this is done, one can construct the eigenvectors, using Eq.~(\ref{eq:transform}). These
  eigenvectors, in turn, can be used to construct the one-particle propagators and to calculate
  the L\"uscher zeta-function in the periodic external field. Since the primary aim
  of the present paper is the proof of principle, we shall not consider all these rather
  straightforward issues here, which form a separate piece of work for the future.

\section{Numerical implementation}
\label{sec:num}

\begin{figure}[t]
  \begin{center}
    \includegraphics*[width=7.cm]{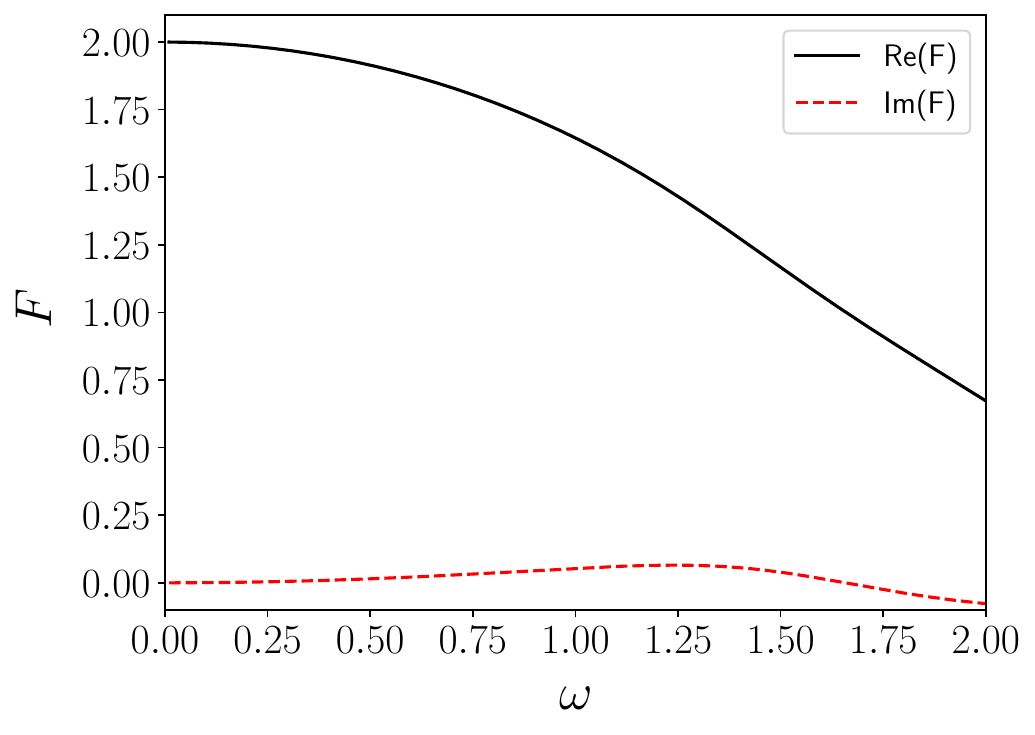}
    \caption{Real and imaginary parts of the resonance form factor.}
    \label{fig:formfactor}
  \end{center}
  \end{figure}

In this section, we shall test our theoretical predictions numerically. Since this test serves
an illustrative purpose only, we
have not made an attempt to choose realistic values of the different parameters in the toy
model. In particular,
we choose $m=1$ from the beginning
and show everything in mass units. The values of other parameters
are $a=-1.5$, $r=-9$, $\kappa=10$ and $C_R=0.9$. Without loss of generality,
one may set $A_0=1$. With this choice of parameters, there exist a couple of poles
on the second Riemann sheet located at $q_R^2=0.123\pm i\,0.082$.
The resonance form factor, evaluated with the help of Eqs.~(\ref{eq:baritGamma})
and (\ref{eq:resFF}), is shown in Fig.~\ref{fig:formfactor}. Note that, owing to the Ward
identity, the form factor is normalized as $F(0)=2$ at $\omega=0$.

\begin{figure}[t]
  \begin{center}
    \includegraphics*[width=0.49\linewidth]{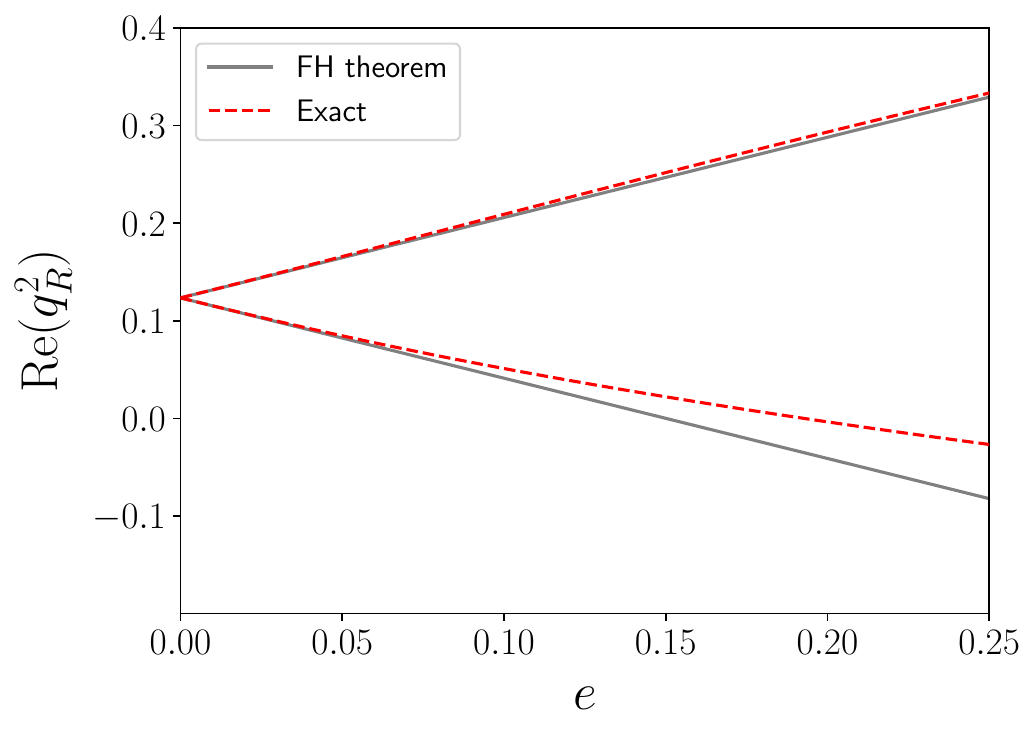}
    \includegraphics*[width=0.49\linewidth]{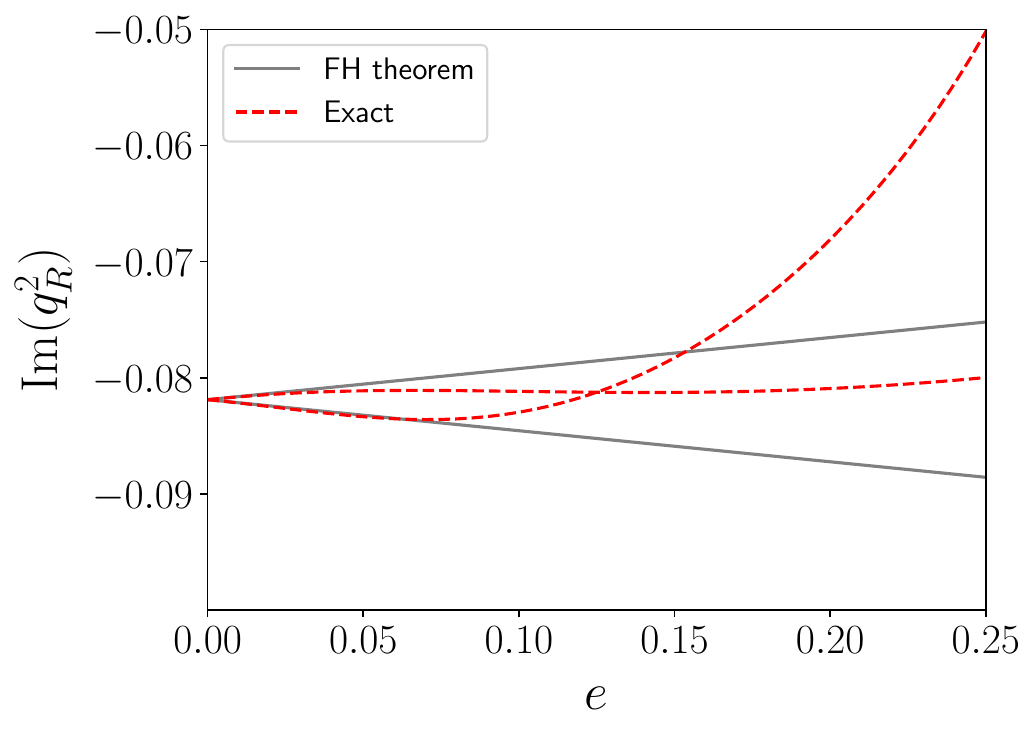}
    \caption{Verification of the Feynman-Hellmann theorem for the real and imaginary parts of the 
      pole position in the complex plane at $\omega=1$.
      Thin black lines depict the prediction of the theorem.}
    \label{fig:FH}
  \end{center}
  \end{figure}

  Furthermore, when $e\neq  0$, the pole
  in the complex plane splits into two that move in the opposite direction from the
  initial location. In Fig.~\ref{fig:FH} we plot the real and imaginary parts of these poles
  versus $e$. It is seen that at small values of $e$ this dependence is almost linear and is
  determined by the Feynman-Hellmann theorem. For this example with $\omega=1$, we obtain
  \eq\label{eq:check-FH}
F({\bf P},{\bf Q}) = 1.6454+i0.0535\, ,\quad\quad
  2  \,\frac{d P_R^0}{d e}\biggr\rvert_{e=0}=1.6455+i0.0534\, .
  \en
  The second number has been obtained by numerically differentiating the pole trajectory in the
  complex plane. The explicit expression of the form factor in this model is written down in
  Appendix~\ref{app:formfactor}.

\begin{figure}[t]
  \begin{center}
    \includegraphics*[width=9cm]{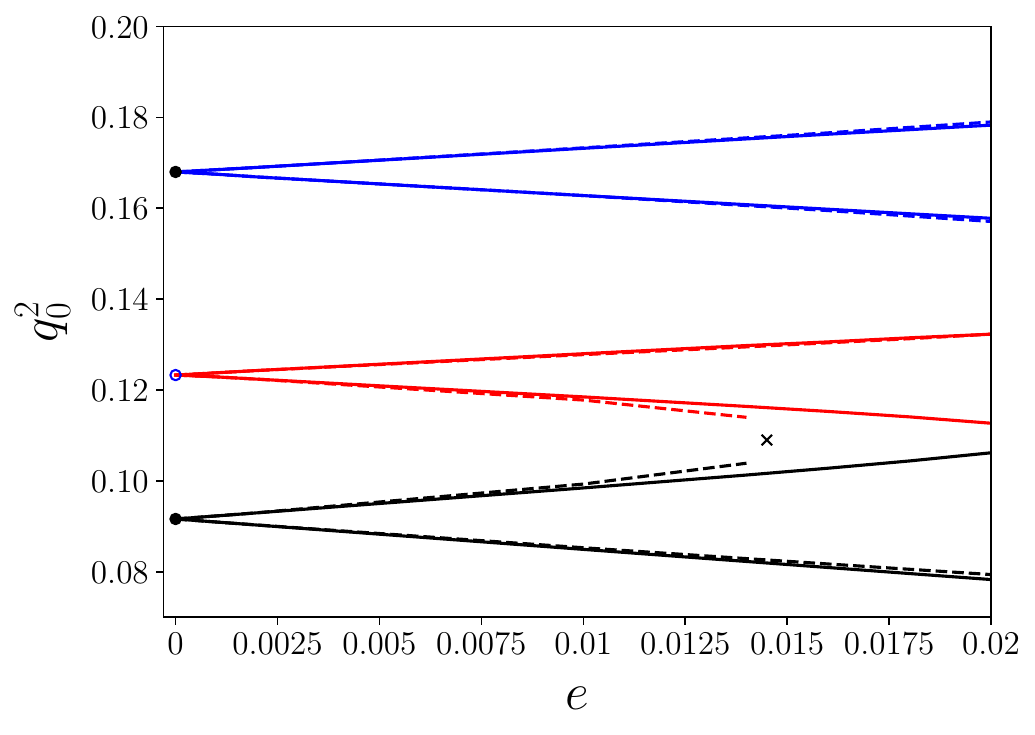}
    \caption{Qualitative structure of the energy levels in the external field. The
      levels at $e=0$ are denoted by black filled dots. These are split when $e\neq 0$.
      There are additional levels (red curves)
      that do not have counterparts at $e=0$. On the figure, they
      emanate from the empty blue dot.
      The solid and dotted lines denote the ``exact''
      and ``perturbative'' solutions, respectively, depending on the use of the ``exact''
      and ``perturbative'' expressions for the loop function. The approximate location where
      two ``perturbative'' levels merge and disappear is marked by a cross.}
    \label{fig:levels}
  \end{center}
  \end{figure}

In Fig.~\ref{fig:levels}, we display the spectrum in a finite volume at different
  values of $e$ and for a fixed $L$ (In order to discuss the qualitative behavior
  of the spectrum we used an arbitrarily chosen value $L=20$).
 The structure of the levels turns out to be
  rather complicated. 
In the absence of field, there is a set of doubly degenerate energy levels (black filled dots) corresponding to states with momentum ${\bf P}$ and $-{\bf P}$, which are related by a time-reversal transformation.  When $e\neq 0$, time-reversal invariance is broken and these two levels split symmetrically at $O(e)$. 
Moreover, there are additional
  energy levels which do not have a zero-field counterpart. This is attributed to the fact that,
  at $e\neq 0$, the poles in the functions ${\it\Pi}_0$, ${\it\Pi}_1$ also split
  (see Fig.~\ref{fig:Pi}), and the determinant in the L\"uscher equation can cross the
  real axis at more places, e.g., between the poles.
  However, it can be easily checked that these solutions correspond to the ``artifacts''
  that were discussed in the previous section. Namely, the residua corresponding to these
  levels are of order $e^2$ and have a different sign as compared to the
  physical levels. We have also checked that the unphysical levels, in
  difference with the physical ones, are not stable
  if the dimension of the matrix in the quantization condition is increased.
  This fact further supports the conclusion that these levels emerge due to
  the approximations that were made during the derivation of the quantization
  condition. Hence, in the analysis of data, such unphysical levels should be merely  discarded.

\begin{figure}[t]
  \begin{center}
    \includegraphics*[width=0.49\linewidth]{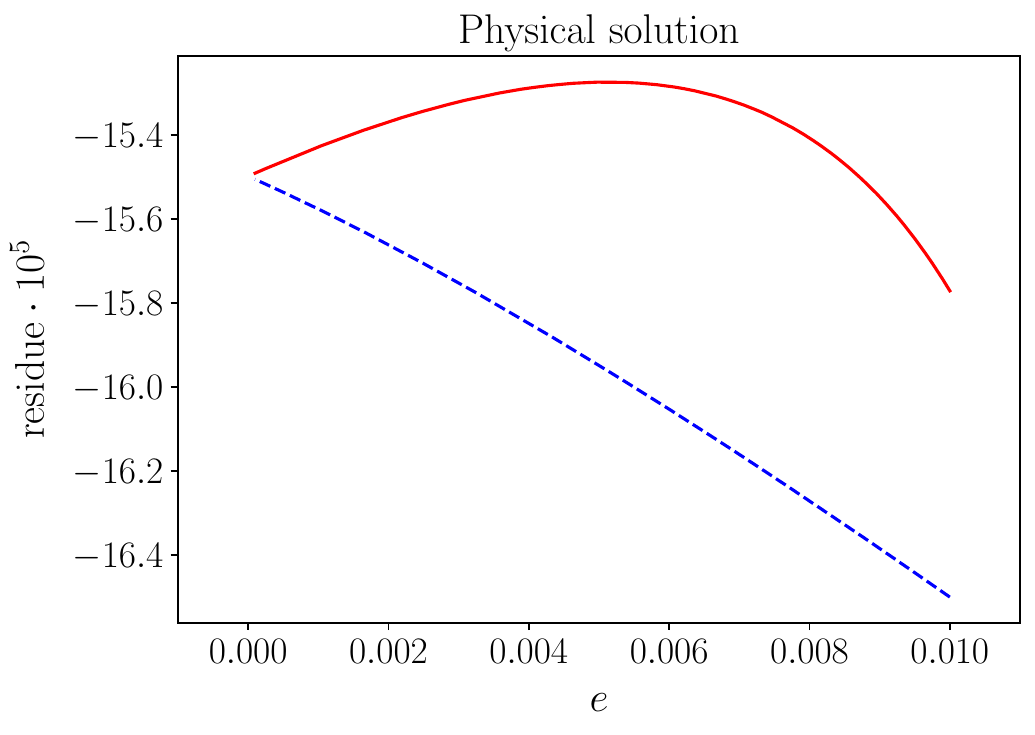}
    \includegraphics*[width=0.49\linewidth]{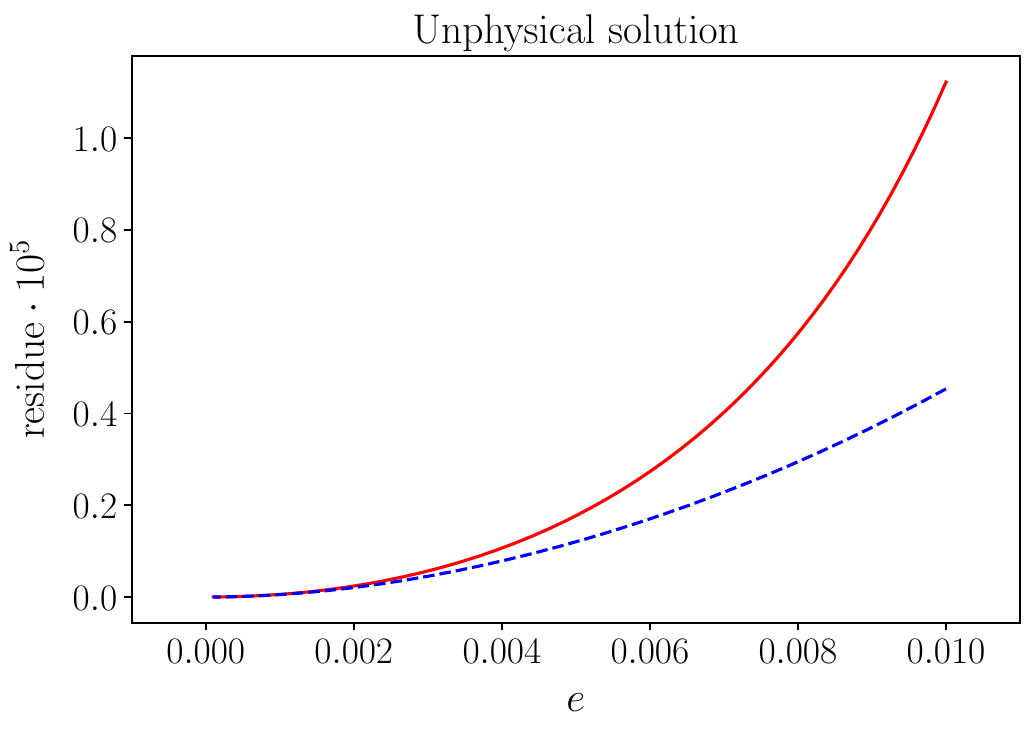}
    \caption{The $e$-dependence of the
      residua for the physical and unphysical levels. The blue and red curves
    correspond to the two different solutions of the quantization condition.}
    \label{fig:residua}
  \end{center}
  \end{figure}

\begin{figure}[t]
  \begin{center}
    \includegraphics*[width=9.cm]{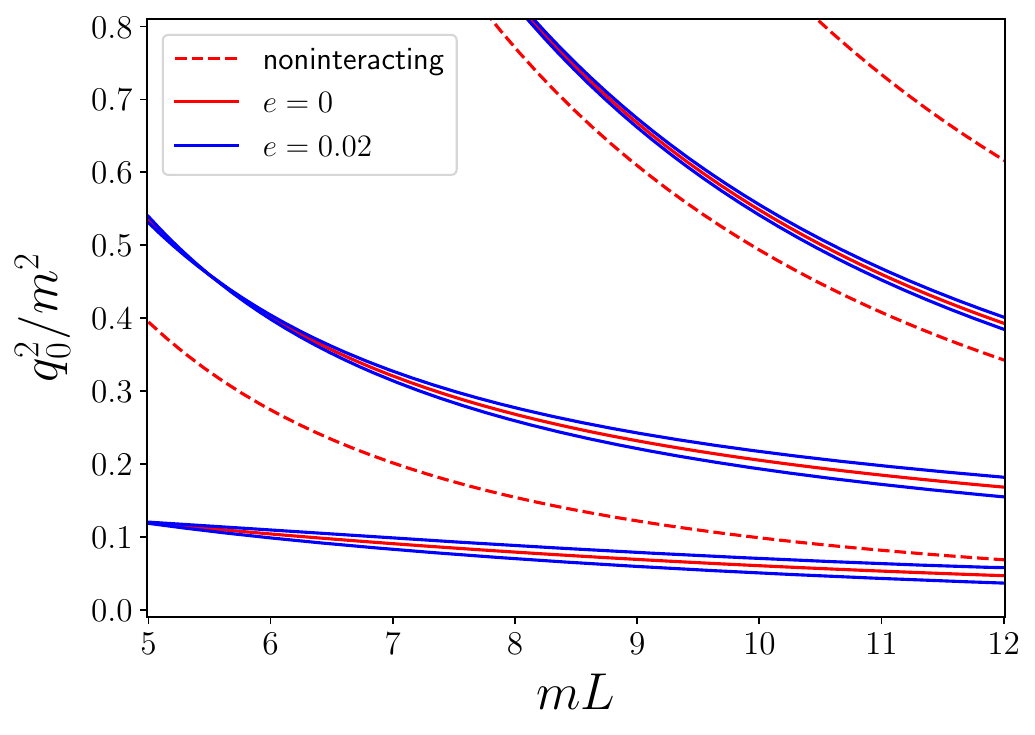}
    \caption{The $L$-dependence of the energy levels for $e=0$, as well as for $e\neq 0$.
      For comparison, we plot the non-interacting energy levels as well, corresponding to the two free particles in a box. A single energy level at $e=0$ splits into two nearby levels,
when the external field is turned on.}
    \label{fig:Ldependence}
  \end{center}
  \end{figure}

  Furthermore,  Fig.~\ref{fig:levels} nicely demonstrates the limitations of the use of
  the  perturbative approach to the calculation of ${\it\Pi}_0$, ${\it\Pi}_1$. In the vicinity
  of $e=0.015$, two ``perturbative'' levels merge and disappear (the determinant
  does not cross the real axis anymore), whereas the ``exact'' levels still exist. Note
  that this happens already at rather small values of $e$, for which other levels are very well
  described by the perturbative solution.

In Fig.~\ref{fig:residua}, the difference between the physical and unphysical levels is clearly
seen. Here, we plot the $e$-dependence of
the residua, calculated using Eq.~(\ref{eq:Zn-1})
(blue and red lines correspond to the two roots of the quantization condition
that merge in the limit $e=0$). For the physical levels, the residua converge to a non-zero
limit and exhibit a linear dependence on $e$ for small $e$. In contrast
to this, the residua for the unphysical levels behave as $e^2$ and vanish
for $e=0$. This agrees with our theoretical findings.

In addition to this, in Fig.~\ref{fig:Ldependence} we show the $L$-dependence of the energy levels
(``exact'' solutions only). 
  Finally, note that, with our choice of parameters, the contribution from the contact
  interaction, parameterized by the coupling $\kappa$, is negligibly small. This could be
  expected, since the first term in Eq.~(\ref{eq:kappa-r}) is much smaller that the second.
  However, as already discussed, this cannot pose an obstacle for the calculation of the
  resonance form factor, the goal we are after.

\section{Conclusions}
\label{sec:concl}

\begin{itemize}

\item[i)]  
  A novel method for the computation of the resonance form factors on the lattice
  has been proposed.
  Within this approach, one circumvents the calculation of the three-point function
  on the lattice, measuring instead the finite-volume energy spectrum in an external periodic field in space. 

\item[ii)]
  It is known that the finite-volume  three-point function, Fig.~\ref{fig:triangle}a, has
  an irregular dependence of the box size $L$ that complicates
  the extraction of the infinite-volume form factor considerably. On the other hand, with this method,
  one merely needs to extract the parameters of the contact interaction
  with the external field from the fit to the energy levels at $e\neq 0$. 
  These parameters, by definition, can only contain exponentially suppressed corrections
  in $L$.
\item[iii)]
  If the lattice simulations are performed at a non-zero external
  field, the formalism that is used to analyze the data should
  be also set in the presence of the external field. In order to
  match this objective,
  a generalization of the L\"uscher equation in the presence of an external periodic field is obtained in this paper.

  \item[iv)]
  Since in the vicinity of the free-particle poles the use of perturbation theory is questionable,
  an expression for the modified L\"uscher function has been derived that avoids the expansion of
  the energy denominators. The limits of the use of perturbation theory
  in this context have been discussed in detail. On the other hand,
  for consistency, one is forced to use a strictly perturbative framework
  to to analyze data on the matrix elements in the ``standard'' approach~\cite{Hoja:2010fm,Bernard:2012bi,Briceno:2015tza,Baroni:2018iau}.
  Of course, in the limit $e\to 0$, the matrix elements extracted within two
  different approaches, agree.
\item[v)]
  The Feynman-Hellmann theorem, which has been so far proven for stable particles only,
  is generalized to the case of resonances. It has been demonstrated that,
  finding the (complex) resonance pole position in the external field and in the Breit frame,
  and differentiating this
  quantity with respect to $e$, one arrives at the form factor.
  The theoretical arguments have been checked numerically, see Eq.~(\ref{eq:check-FH}).

\item[vi)]
  A numerical implementation of the framework is considered for a toy model.
    The qualitative structure of the energy levels is discussed.

  \item[vii)]
    The present work provides a proof of principle only. Different improvements and
    generalizations will have to be considered. For example, it will be crucial
    to take into account  relativistic corrections to all orders and write down a
    framework that is explicitly Lorentz-invariant. Furthermore, higher orders in the effective
    theory should be systematically included in order to write down the result
    in a form that does not explicitly rely on the effective-range expansion, and is also valid
    away from the elastic threshold. Partial-wave mixing should also be addressed
    appropriately. Finally, the numerical implementation should be considered
    for realistic values of the parameters that resemble the cases of existing low-lying
    resonances. All these technical issues will be addressed in the future.

\end{itemize}

\begin{acknowledgments}

We thank A. Wirzba for useful discussions. JL would like to acknowledge the financial support from the fellowship
``Regierungsstipendiaten CONACYT-DAAD mit Mexiko''  under the grant number 2016 (57265507).
The work of UGM and AR was funded in part by the Deutsche Forschungsgemeinschaft
(DFG, German Research Foundation) – Project-ID 196253076 – TRR 110,
Volkswagenstiftung (grant no. 93562), and the Chinese Academy of Sciences (CAS) President's
International Fellowship Initiative (PIFI) (grant nos. 2021VMB0007 and 2018DM0034).
The work of FRL has been supported by the U.S. Department of Energy, Office of Science,
Office of Nuclear Physics, under grant Contract Numbers DE-SC0011090 and DE-SC0021006.

\end{acknowledgments}

\appendix

\section{Mathieu equation: essentials}
\label{app:Mathieu}

The differential equation~(\ref{eq:differential}) admits variable separation by using the ansatz
\eq\label{eq:separation}
\Phi({\bf x},t)=e^{-iEt+i{\bf p}_\perp{\bf x}_\perp}\bar\Phi(z)\, ,\quad\quad
z=\frac{\omega x_\parallel}{2}\, ,
\en
where the function $\bar\Phi(z)$ obeys the differential equation
\eq\label{eq:Mathieu1}
\biggl(\frac{d^2}{dz^2}+\frac{8m}{\omega^2}\biggl(E-m-\frac{{\bf p}_\perp^2}{2m}\biggr)
+\frac{8me{\it\Gamma}A_0}{\omega^2}\,
\cos 2z\biggr)\bar\Phi(z)=0\, .
\en
This coincides with the Mathieu equation~\cite{McLachlan,NIST}.
Note that the potential in the above equation is periodic, and hence
the solutions are given by Bloch's wave functions that have the
property
\eq
\bar\Phi(z+\pi)=e^{i\nu\pi}\bar\Phi(z)\, , \quad\quad
-1<\nu\leq 1\, .
\en
The solutions corresponding to a particular $\nu$ (the so-called $\nu$-periodic solutions)
are denoted by ${\rm me}_{\nu+2n}(z,q)$ (with an integer $n$) and obey the equation
\eq\label{eq:Mathieu2}
\biggl(\frac{d^2}{dz^2}+\lambda_{\nu+2n}(q)-2q\cos 2z\biggr){\rm me}_{\nu+2n}(z,q)=0\, .
\en
From the comparison of Eqs.~(\ref{eq:Mathieu1}) and (\ref{eq:Mathieu2}) it follows that
\eq\label{eq:q-A0}
\lambda_{\nu+2n}(q)=\frac{8m}{\omega^2}\biggl(E-m-\frac{{\bf p}_\perp^2}{2m}\biggr)\, .
\en
In case when $\nu$ becomes integer, one has
\eq
\lambda_n(q)=\left\{
  \begin{array}{l l}
    a_n(q)\,,&\quad n=0,1,\cdots\cr
               b_{-n}(q)\,,&\quad n=-1,-2,\cdots
  \end{array}\right.
\en
and
\eq
me_n(z,q)=\left\{
  \begin{array}{ll}
    \sqrt{2}\,ce_n(z,q), &\quad n=0,1,\cdots
    \\[2mm]
    -\sqrt{2}i\,se_{-n}(z,q),&\quad n=-1,-2,\cdots
  \end{array}
\right.
\en
Due to the periodic boundary conditions, the parameter $\nu$ will be
quantized. Indeed, from $\Phi(x_\parallel+L)=\Phi(x_\parallel)$ we get
$\bar\Phi(z+\pi N)=e^{i\nu\pi N}\bar\Phi(z)=\bar\Phi(z)$, leading
to the condition $e^{i\nu\pi N}=1$. Together with the requirement
$-1< \nu\leq 1$ this leads to the conclusion that $\nu$ can take the following values
\eq\label{eq:nui}
  \begin{array}{lll}
    N=1: &\quad& \nu=0 \\[2mm]
    N=2: &\quad& \nu=0,1 \\[2mm]
    N=3: &\quad& \nu=-\frac{2}{3},0,\frac{2}{3} \\[2mm]
    N=4: &\quad& \nu=-\frac{1}{2},0,\frac{1}{2},1 \\[2mm]
  \end{array}
\en
and so on.

The Fourier expansion of the Mathieu functions takes the form
\eq\label{eq:Fourier}
{\rm me}_\nu(z,q)=\sum_{a=-\infty}^\infty C_{2a}^\nu(q)
e^{i(\nu+2a)z}\, .
\en
The coefficients of this expansion, $C_{2a}^\nu(q)$, are known~\cite{NIST}.

\section{Expansion of the propagator in powers of $q$}
\label{app:expansion}

The expansion of Mathieu functions $me_\nu(z,q)$ in powers of $q$
for the non-integer ($\nu$) and integer ($k\geq 2$) values of the index is
given by
\eq\label{eq:menu_q}
{\rm me}_\nu(z,q)&=&e^{i\nu z}-\frac{q}{4}\,\biggl(\frac{1}{\nu+1}\,e^{i(\nu+2)z}
-\frac{1}{\nu-1}\,e^{i(\nu-2)z}\biggr)+O(q^2)\, ,
\nonumber\\[2mm]
{\rm me}_k(z,q)&=&\sqrt{2}\biggl\{\cos kz-\frac{q}{4}\,\biggl(\frac{1}{k+1}\,\cos(k+2)z
-\frac{1}{k-1}\,\cos(k-2)z\biggr)+O(q^2)\biggr\}\, ,
\nonumber\\[2mm]
{\rm me}_{-k}(z,q)&=&-i\sqrt{2}\biggl\{\sin kz-\frac{q}{4}\,\biggl(\frac{1}{k+1}\,\sin(k+2)z
-\frac{1}{k-1}\,\sin(k-2)z\biggr)+O(q^2)\biggr\}\, .
\nonumber\\
&&
\en
If $k=1,0,-1$, the pertinent expressions take the form
\eq\label{eq:me1_q}
{\rm me}_1(z,q)&=&\sqrt{2}\biggl\{\cos z-\frac{q}{8}\,\cos 3z+O(q^2)\biggr\}\, ,
\nonumber\\[2mm]
{\rm me}_0(z,q)&=&1-\frac{q}{2}\,\cos 2z+O(q^2)\, ,
\nonumber\\[2mm]
{\rm me}_{-1}(z,q)&=&-i\sqrt{2}\biggl\{\sin z-\frac{q}{8}\,\sin 3z+O(q^2)\biggr\}\, .
\en
In order to perform the expansion of
the two-point function $S({\bf x},{\bf y };E)$, given by Eq.~(\ref{eq:SA_x}),
one should consider the cases of odd and even $N$ separately.
In case of the odd $N$, the only
integer value of the parameter $\nu$ in the interval $\nu\in ]-1,1]$ is $\nu=0$.
In case of the even $N$, there are two integer values $\nu=0,1$. In the sum over
all eigenvectors, one should separate the integer and non-integer values of $\nu$,
and carry out the expansion in $q$ in each term.

Let us start from the more simple case of the odd $N$. Here, $i=1$ corresponds to the
value $\nu_i=0$. The eigenvalues are given by $\lambda_{\nu_i+2n}=(\nu_i+2m)^2+O(q^2)$.
The original expression of the propagator can be split into three terms
$S= S_1+S_2+S_3$, where, at $O(q^2)$,
\eq\label{eq:S1}
S_1 &=& \frac{1}{L^3} \sum_{{\bf p}_\perp} \sum_{i=2}^N \sum_{n=-\infty}^\infty \frac{e^{i {\bf p}_\perp ({\bf x}_\perp - {\bf y}_\perp)}}{m + \dfrac{{\bf p}_\perp^2}{2m} + \dfrac{\omega^2}{8m}(\nu_i +2n)^2  - E} \nonumber\\[2mm]
& \times& \left[  e^{i (\nu_i+2n) \frac{\omega  x_\parallel}{2}} -\frac{q}{4}
  \left( \frac{e^{i(\nu_i+2n+2) \frac{\omega  x_\parallel}{2}}}{\nu_i+2n+1}
    - \frac{e^{i(\nu_i+2n-2)\frac{\omega  x_\parallel}{2}}}{\nu_i+2n-1}
     \right) \right] \nonumber\\[2mm]
   & \times& \left[  e^{-i(\nu_i+2n) \frac{\omega y_\parallel}{2}} -\frac{q}{4}
     \left( \frac{e^{-i(\nu_i+2n+2)\frac{\omega  y_\parallel}{2}}}{\nu_i+2n+1}
       - \frac{e^{-i(\nu_i+2n-2)\frac{\omega  y_\parallel}{2}}}{\nu_i+2n-1}
         \right)  \right]\, ,
\nonumber\\[2mm]
&=& \frac{1}{L^3} \sum_{{\bf p}_\perp}
\sum_{i=2}^N \sum_{n=-\infty}^\infty
\frac{e^{i {\bf p}_\perp ({\bf x}_\perp - {\bf y}_\perp)}}
{m + \dfrac{{\bf p}_\perp^2}{2m} + \dfrac{\omega^2}{8m}(\nu_i +2n)^2  - E}\,
e^{i (\nu_i+2n) \frac{\omega  (x_\parallel-y_\parallel)}{2}}
\nonumber\\[2mm]
& \times&
 \left[  1 -\frac{q}{4}\,
   \left( \frac{e^{i\omega  x_\parallel}}{\nu_i+2n+1}
     - \frac{e^{-i\omega  x_\parallel} }{\nu_i+2n-1}
     + \frac{e^{-i\omega  y_\parallel}}{\nu_i+2n+1}
     - \frac{e^{i\omega  y_\parallel} }{\nu_i+2n-1}  \right)  \right]\, ,
  \en
\eq
S_2 = \frac{1}{L^3} \sum_{{\bf p}_\perp} \sum_{n=0}^\infty
\frac{e^{i {\bf p}_\perp ({\bf x}_\perp - {\bf y}_\perp)}}{m + \dfrac{{\bf p}_\perp^2}{2m} + \dfrac{\omega^2}{8m}(2n)^2 - E} \left[2 {\rm ce}_{2n}\biggl(\frac{\omega x_\parallel}{2},q\biggr)
  {\rm ce}_{2n}\biggl(-\frac{\omega y_\parallel}{2},q\biggr) \right]\,,
\en
and
\eq
S_3 = \frac{1}{L^3} \sum_{{\bf p}_\perp} \sum_{n=-\infty}^{-1}
\frac{e^{i {\bf p}_\perp ({\bf x}_\perp - {\bf y}_\perp)}}{m + \dfrac{p_\perp^2}{2m} + \dfrac{\omega^2}{8m}(2n)^2  - E} \left[- 2 {\rm se}_{2n}\biggl(\frac{\omega x_\parallel}{2},q\biggr)
  {\rm se}_{2n}\biggl(-\frac{\omega y_\parallel}{2},q\biggr) \right]\,.
\en
Using Eq.~(\ref{eq:me1_q}), one could rewrite the last two terms at $O(q)$ in the following form:
\eq\label{eq:S23}
S_2 + S_3&=&
=\sum^{\infty}_{n=-\infty}
\sum_{{\bf p}_\perp}
\frac{e^{i {\bf p}_\perp ({\bf x}_\perp - {\bf y}_\perp) }  }
{m + \dfrac{{\bf p}_\perp^2}{2m} +\dfrac{\omega^2 n^2}{2m} - E}\,
e^{in\omega (x_\parallel - y_\parallel)}
\nonumber\\[2mm]
&\times&\Biggl\{1-\frac{q}{4}\left(\frac{e^{-i\omega y_\parallel}}{2n+1}-\frac{e^{i\omega y_\parallel}}{2n-1}+\frac{e^{i\omega x_\parallel}}{2n+1}-\frac{e^{-i\omega x_\parallel}}{2n-1}\right)\Biggr\}\,.
\en
It is easy to see that Eq.~(\ref{eq:S23}) follows from Eq.~(\ref{eq:S1}) for $\nu_i=0$. Hence,
one could lump together these two expressions, extending the sum in Eq.~(\ref{eq:S1})
from $i=1$ to $i=N$. Furthermore, defining $p_\parallel=\frac{\omega}{2}\,(\nu_i+2n)$, it
is easily seen that the sum over all $i,n$ is equivalent to sum over all $p_\parallel=\frac{2\pi}{L}\,k$,
where $k\in\mathbb{Z}$. Defining further ${\bf p}=({\bf p}_\perp,p_\parallel)$
and ${\bf p}{\bf x}={\bf p}_\perp{\bf x}_\perp-p_\parallel x_\parallel$, the two-point function
can be rewritten in a more compact form:
\eq
S({\bf x},{\bf y};E)&=& \frac{1}{L^3}\,\sum_{{\bf p}}
\frac{e^{i {\bf p}({\bf x}-{\bf y})}}{m+\dfrac{{\bf p}^2}{2m}-E}
\nonumber\\[2mm]
&\times& \biggl\{1-\frac{\omega q}{8}\left( \frac{e^{i\omega x_\parallel}}{p_\parallel + \frac{\omega}{2}}-\frac{e^{-i\omega x_\parallel}}{p_\parallel - \frac{\omega}{2}}+\frac{e^{-i\omega y_\parallel}}{p_\parallel + \frac{\omega}{2}}-\frac{e^{i\omega y_\parallel}}{p_\parallel - \frac{\omega}{2}}\right)\biggr\}\, .
\en
One can now shift $p_\parallel \to p_\parallel - \omega$ and
$p_\parallel \to p_\parallel + \omega$ in the third and fourth terms in the brackets,
respectively. Then, we have
\eq
&&S({\bf x},{\bf y},E)\,=\, \frac{1}{L^3}\sum_{\bf p}
\frac{e^{i {\bf p}({\bf x}-{\bf y})}}{m+\dfrac{{\bf p}^2}{2m}-E}\nonumber\\[2mm]
&-&\biggl\{
\frac{\omega q}{8}\frac{1}{L^3}\sum_{\bf p}
\frac{e^{i ({\bf p}+\boldsymbol{\omega}){\bf x}-i{\bf p}{\bf y}}}{p_\parallel +\frac{\omega}{2}} \biggl(\frac{1}{m+\dfrac{{\bf p}^2}{2m}-E}
-\frac{1}{m+\dfrac{({\bf p}+\boldsymbol{\omega})^2}{2m}-E}\biggr)
+(\omega \to -\omega)\biggr\}\nonumber\\[2mm]
&=&\frac{1}{L^3}\sum_{\bf p}
\frac{e^{i {\bf p}({\bf x}-{\bf y})}}{m+\dfrac{{\bf p}^2}{2m}-E}
\nonumber\\[2mm]
&-& \frac{\omega^2 q}{8}\frac{1}{L^3}\sum_{\bf p}
\biggl\{
\frac{e^{i({\bf p} + \boldsymbol{\omega}){\bf x}- i {\bf p}{\bf y}}}
{\biggl(m+\dfrac{{\bf p}^2}{2m}-E\biggr)\biggl(m+\dfrac{({\bf p}+\boldsymbol{\omega})^2}{2m}-E\biggr)}
+(\omega \to -\omega)\biggr\}\, . 
\en
Performing the Fourier transform and using Eq.~(\ref{eq:solutions}), we finally arrive at
Eq.~(\ref{eq:perturbative}).

The calculations in case of an even $N$ are slightly more complicated. Now, the eigenvalue
corresponding to $\nu_i=1$ is also present, with
$\lambda_{\pm 1}(q)=1\pm q+O(q^2)$. Hence, the denominators corresponding to this eigenvalue,
should be expanded:
\eq
\frac{1}{m+\dfrac{{\bf p}_\perp^2}{2m}+\dfrac{\omega^2}{8m}\,\lambda_{\pm 1}(q)-E}&=&\frac{1}{m+\dfrac{{\bf p}_\perp^2}{2m}+\dfrac{\omega^2}{8m}-E}
\nonumber\\[2mm]
&\mp&\frac{\omega^2q}{8m}\,\frac{1}{\biggl(m+\dfrac{{\bf p}_\perp^2}{2m}+\dfrac{\omega^2}{8m}-E\biggr)^2}+O(q^2)\, .
\en
Otherwise, the calculations follow exactly the same path. Adding all contributions carefully,
one finally verifies that Eq.~(\ref{eq:perturbative}) holds in case of the even $N$ as well.

\section{The L\"uscher function at $e\neq 0$}
\label{app:Luescher-modified}

Taking into account the fact that $\omega=2\pi N/L$ and performing the variable transformation
$x_\parallel=2u/\omega$, $y_\parallel=2v/\omega$
in Eq.~(\ref{eq:Luescher-modified}), we get
\eq
&&\bar{\it\Pi}(P_\parallel,Q_\parallel;{\bf P}_\perp;E)
=\frac{1}{\pi^2L^2N^2}\,\sum_{{\bf p}_\perp}
\sum_{i,j=1}^N\sum_{n,m=-\infty}^\infty\int_0^{N\pi} du \int_0^{N\pi} dv\,
D_{in,jm}({\bf p}_\perp;{\bf P}_\perp;E)
\nonumber\\[2mm]
&\times&e^{-iau+ibv}\,{\rm me}_{\nu_i+2n}(u,q)
{\rm me}_{\nu_i+2n}(-v,q)
{\rm me}_{\nu_j+2m}(u,q)
{\rm me}_{\nu_j+2m}(-v,q)\, .\quad\quad
\en
Here, $a=2P_\parallel/\omega$  and $b=2Q_\parallel/\omega$. Furthermore,
using the periodicity property of the Mathieu functions, the integration over the variables
$u,v$ can be restricted to the interval from $0$ to $\pi$:
\eq
&&\int_0^{N\pi} du \int_0^{N\pi} dv\,e^{-iau+ibv}\,{\rm me}_{\nu_i+2n}(u,q)
{\rm me}_{\nu_i+2n}(-v,q)
{\rm me}_{\nu_j+2m}(u,q)
{\rm me}_{\nu_j+2m}(-v,q)
\nonumber\\[2mm]
&=&\sum_{k,l=1}^{N-1}e^{i\pi(\nu_i+2n+\nu_j+2m-a)(k-1)-i\pi(\nu_i+2n+\nu_j+2m-b)(l-1)}
\nonumber\\[2mm]
&\times&\int_0^\pi du \int_0^\pi dv\,e^{-iau+ibv}\,{\rm me}_{\nu_i+2n}(u,q)
{\rm me}_{\nu_i+2n}(-v,q)
{\rm me}_{\nu_j+2m}(u,q)
{\rm me}_{\nu_j+2m}(-v,q)
\nonumber\\[2mm]
&=&N^2\sum_{k,l=-\infty}^\infty \delta_{a-b,2k}\,\delta_{\nu_i+2n+\nu_j+2m-a,2l}
\nonumber\\[2mm]
&\times&\int_0^\pi du \int_0^\pi dv\,e^{-iau+ibv}\,{\rm me}_{\nu_i+2n}(u,q)
{\rm me}_{\nu_i+2n}(-v,q)
{\rm me}_{\nu_j+2m}(u,q)
{\rm me}_{\nu_j+2m}(-v,q)\, .
\nonumber\\
&&
\en
Hence,
\eq
\bar{\it\Pi}(P_\parallel,Q_\parallel;{\bf P}_\perp;E)=\sum_{\ell=-\infty}^\infty
L\delta_{P_\parallel-Q_\parallel,\ell\omega}\tilde{\it\Pi}(P_\parallel,Q_\parallel;{\bf P}_\perp;E)\, ,
\en
where
\eq
&&\tilde{\it\Pi}(P_\parallel,Q_\parallel;{\bf P}_\perp;E)=
\frac{1}{\pi^2L^3}\,\sum_{{\bf p}_\perp}
\sum_{i,j=1}^N\sum_{n,m=-\infty}^\infty
\sum_{k=-\infty}^\infty \delta_{\nu_i+2m+\nu_j+2m-a,2k}
\nonumber\\[2mm]
&\times&\int_0^\pi du \int_0^\pi dv\,
D_{in,jm}({\bf p}_\perp;{\bf P}_\perp;E) e^{-iau+ibv}
\nonumber\\[2mm]
&\times&
{\rm me}_{\nu_i+2n}(u,q)
{\rm me}_{\nu_i+2n}(-v,q)
{\rm me}_{\nu_j+2m}(u,q)
{\rm me}_{\nu_j+2m}(-v,q)\, .
\en
Since $\nu_i,\nu_j\in ]-1,1]$, the sum over $k$ in the above equation has a finite number
of non-zero terms. Finally, one can carry out the summation over $k$, which yields
\begin{align}
\begin{split}
\label{eq:Pitilde}
&\tilde{\it\Pi}(P_\parallel,Q_\parallel;{\bf P}_\perp;E)=
\frac{1}{\pi^2L^3}\,\sum_{{\bf p}_\perp}
\sum_{i,j=1}^N\sum_{n,m=-\infty}^\infty
\int_0^\pi du \int_0^\pi dv\,
D_{in,jm}({\bf p}_\perp;{\bf P}_\perp;E) e^{-iau+ibv}
\\
&\times
{\rm me}_{\nu_i+2n}(u,q)
{\rm me}_{\nu_i+2n}(-v,q)
{\rm me}_{\nu_j+2m}(u,q)
{\rm me}_{\nu_j+2m}(-v,q)\, .
\end{split}
\end{align}
Note that the conservation of the ``longitudinal momentum'' takes the form
\eq
\frac{\omega}{2}\,(\nu_i+2n)+\frac{\omega}{2}\,(\nu_j+2m)-P_\parallel=k\omega\, .
\en
Equation~(\ref{eq:Pitilde}) is still too complicated for using it in the analysis of data.
Here, we are interested in the shift of the energy levels that are linear in $e$. 
It would be therefore  useful to get a simplified expression that allows one to
extract the levels at this precision.
To this end, one first expands the numerator, using the Eqs.~(\ref{eq:menu_q}) and (\ref{eq:me1_q}).
Furthermore, as we already know, the eigenvalues $\lambda_{\nu_i+2n}(q)$ up to the
order $q^2$ correspond to those in the free theory,
whereas the case $\nu_i+2n=\pm 1$ is an exception, see Eq.~(\ref{eq:lambda1-1}).
Expanding the denominator
in $D_{in,jm}({\bf p}_\perp;{\bf P}_\perp;E)$ up to the first order in $q$
corresponds to the ``perturbative'' expression, whereas
leaving the denominator intact leads to the ``exact'' one. The final result
is displayed in Eqs.~(\ref{eq:Piini}), (\ref{eq:Piprime}), (\ref{eq:Pi2prime}).

\section{Explicit expression for the form factor}
\label{app:formfactor}

An explicit expression for the form factor in the toy model considered here
can be straightforwardly obtained by evaluating the expression
given in Eq.~(\ref{eq:resFF}). Below, we give the final result without derivation:
\eq
F({\bf P},{\bf Q})=F(\omega)=\frac{\sqrt{-q_R^2}}{4\pi\left(1+r\sqrt{-q_R^2}\right)}\,
\biggl\{-\kappa\omega^2q_R^2+8\pi{\it\Gamma}\biggl(r+\frac{4}{\omega}\,
\arcsin\frac{\omega}{\sqrt{\omega^2-16q_R^2}}\biggr)\biggr\}\, .
\nonumber\\
\en

\end{document}